\documentclass[12pt,preprint]{aastex}
\shorttitle{2-D stellar evolution code}
\newcommand{\p}[2]{\frac{\partial #1}{\partial #2}}

\newcommand{\od}[2]{\frac{d #1}{d #2}}
\newcommand{\s}[1]{\mbox{\scriptsize{#1}}}
\newcommand{\vb}[1]{{\bf #1}}

\newcommand{\h}[1]{{\cal#1}}

\newcommand{\be}{\begin{equation}}
\newcommand{\ee}{\end{equation}}
\newcommand{\ba}{\begin{eqnarray}}
\newcommand{\ea}{\end{eqnarray}}
\newcommand{\Ba}{\begin{eqnarray*}}
\newcommand{\Ea}{\end{eqnarray*}}

\newcommand{\pa}[2]{\left(\frac{\partial #1}{\partial #2}\right)_m}

\newcommand{\unit}[1]{\hat{\bf e}_{#1}}

\begin{document}

\title{TWO-DIMENSIONAL STELLAR EVOLUTION CODE INCLUDING ARBITRARY MAGNETIC FIELDS. II. 
PRECISION IMPROVEMENT AND INCLUSION OF TURBULENCE AND ROTATION}

\author{Linghuai Li \altaffilmark{1}, Sabatino Sofia\altaffilmark{1}, Paolo Ventura\altaffilmark{2}, 
Valentina Penza\altaffilmark{3}, Shaolan Bi\altaffilmark{4}, Sarbani Basu\altaffilmark{1}, 
and Pierre Demarque\altaffilmark{1}}
\altaffiltext{1}{Department of Astronomy, Yale University, P.O. Box 208101, New Haven, CT 06520-8101}
\email{li@astro.yale.edu} 
\altaffiltext{2}{INAF, Osservatorio Astronomico di Roma, 00040 Monteporzio Catone (RM), Italy}
\altaffiltext{3}{Universita' Di Roma "Tor Vergata", Via Della Ricerca Scientifica 1, 00133 Roma, Italy}
\altaffiltext{4}{Department of Astronomy, Beijing Normal University, Beijing 100875, China}

\begin{abstract}
In the second paper of this series we pursue two objectives. First, in 
order to make the code more sensitive to small effects, we remove many
approximations made in Paper I. Second, we include turbulence and rotation 
in the two-dimensional framework. The stellar equilibrium is described by means
of a set of five differential equations, with the introduction of a new dependent
variable, namely the perturbation to the radial gravity, that is found when 
the non-radial effects are considered in the solution of the Poisson
equation; following the scheme of the first paper, we write the equations
in such a way that the two-dimensional effects can be easily disentangled. 
The key concept introduced in this series is the equipotential surface. We use
the underlying cause-effect relation to develop a recurrence relation to
calculate the equipotential surface functions for uniform rotation, differential
rotation, rotation-like toroidal magnetic fields and turbulence.
We also develop a more precise code to numerically solve the
two-dimensional stellar structure and evolution equations based on the
equipotential surface calculations. We have shown that with this formulation we can 
achieve the precision required by observations by appropriately selecting the 
convergence criterion. Several examples are presented to show 
that the method works well. Since we are interested in modeling the effects of a dynamo-type 
field on the detailed envelope structure and global properties of 
the Sun, the code has been optimized for short timescales phenomena 
(down to 1 yr).  The time dependence of the code has so far
been tested exclusively to address such problems.
\end{abstract}

\section{INTRODUCTION}\label{sect:s1}

High precision is an essential requirement in solar variability modeling
because the cyclical variations of all solar global parameters are very
small (see Li et al 2003 and references therein). For example, the (relative) precision 
of the measurements of the TSI is about $10^{-5}$. Oscillation splittings can also be 
measured with a similar precision, and the PICARD satellite expects to measure 
diameter changes with a precision of a few milli-arc seconds, thus a few parts in $10^6$.
These requirements are even more extreme in the two-dimensional (2D) case, 
because two-dimensional effects are subtler than their 1D counterparts.
This gives us a sense of the precision required for our code.

In the first paper of this series  (Li et al 2006, referred hereafter as
Paper I), we developed a 2D stellar evolution code that includes magnetic
fields of arbitrary cylindrically symmetric configuration by generalizing
in a straightforward way our one-dimensional (1D) code (Lydon \&\ Sofia
1995; Li \&\ Sofia 2001; Li et al 2002; Li et al 2003). Since the 2D
case is very complex, we made some significant approximations, some
physical, and some computational. In terms of the physical approximations,
for the first two, we neglected the second-order derivative of the
gravitational potential $\Phi$ with respect to the colatitude coordinate
$\theta$ and the second-order derivative of the perturbation
gravitational potential $\Phi-\Phi_0$ with respect to the radial
coordinate $r$, where $\partial\Phi_0/\partial r=Gm/r^2$ is the
spherically-symmetric gravitation acceleration component in the radial
direction, i.e., expression (30) in Paper I. The third approximation is
that we ignored turbulence, which had been included in our 1D variability
models (Li et al 2002). A detailed comparison of the 1D solar 
variability models with the relevant observations (Li et al 2003) shows
that turbulence must play an important role. In particular, in order to
explain the changes of the oscillation spectrum in function of the
activity cycle, we needed to include a model of turbulence that interacts
with magnetic fields in a negative feedback sense.  In this paper we
remove these three physical approximations made in Paper I.

Unlike the three approximations mentioned above, the fourth approximation
made in Paper I is computational, involving the solution method of the 2D
stellar structure equations. In the 1D case, we use the trapezoidal rule
to integrate the 1D stellar structure equations. In the 2D version in
Paper I, the trapezoidal rule (or the central difference scheme) was not
applied everywhere, since we used numerical derivatives. In this paper we
minimize the use of numerical derivatives. The fifth approximation made
in Paper I is that we neglected $\partial F_\theta/\partial\theta$ in the
luminosity equation, i.e., the term $O(2)$ in Eq.~(124d), which we now
include. The similar term in Eq.~(124e) of Paper I does not matter for
the cyclic variation of the Sun. 

Removal of the above six approximations is one of the main objectives of this
paper. The second main objective is to include turbulence and rotation, which are
also important sources for asphericity. In \S\ref{sct:s2} we summarize the
theoretical foundations that give rise to the 2D stellar variability models by
including magnetic fields, turbulence and rotation. Since we want to get rid 
of approximations 1 and 2, we have to add the Poisson equation (which is a
second-order partial differential equation) to the stellar structure equations.
We thus have two more first-order stellar structure equations. As a result, we
now have a total of six stellar structure equations.

Equipotential surface is the key concept to obtain the 2D generalization from the
1D stellar structure and evolution equations. In \S\ref{sct:s3} we show how 
to find out the equipotential surface from the 2D stellar structure equations
obtained in \S\ref{sct:s2}. Magnetic fields, turbulence and rotation are causes,
and the resultant matter redistribution is the effect. This cause-effect relation
indicates certain recurrence relation for equipotential surface calculations. We
present the recurrence relations for the uniform rotation, differential rotations,
rotation-like toroidal magnetic fields and turbulence in \S\ref{sct:s3}.

The third main objective of this paper is to raise the numerical precision 
of the numerical solutions for the 2D stellar structure and evolution 
equations. We tried hard to do so and found out that it is the best to 
explicitly invoke the equipotential surface. We present this method of solution
in \S\ref{sct:s4}. We give a typical example 
of 2D solar variability models in \S\ref{sct:s5} to show how we use 
this 2D code. The conclusion is presented in the last section.

\section{THEORETICAL FOUNDATION}\label{sct:s2}

Magnetic fields, turbulence, and rotation are possible 
causes of asphericity. In this paper, we consider all of them. 
We assume that the system is azimuthally symmetric 
or axisymmetric. Therefore, we need only the radius ($r$) and 
colatitude ($\theta$) in the spherical polar coordinate 
($r,\theta,\phi$), in which the azimuthal angle $\phi$ is 
irrelevant. The basic equations represent the conservation of 
mass, momentum, and energy. We also need the Poisson equation and 
the energy transport equation to close the system. Since magnetic 
fields are involved, the Maxwell equations must also be obeyed, 
for example, we require $\nabla\cdot\vb{B}=0$. In this 
section we summarize the results and point out the differences 
from their 1D counterpart.

\subsection{Mass Conservation}

Mass conservation is guaranteed by calculating the mass enclosed 
within a certain surface. In a spherically symmetric system the 
surface is a spherical surface with radius $r$ with respect to the 
symmetric center of the system. This spherical surface is also an 
equipotential surface of gravity. In the general case the 
equipotential surface $r=R(\Phi,\theta)$ is thus used to define the 
mass $M_\Phi$:
\begin{equation}
  M_\Phi = 2\pi\int_0^\pi d\theta\sin\theta\int_0^{R(\Phi,\theta)}
   dr r^2\rho(r,\theta). \label{eq:mass1}
\end{equation}
In the spherically symmetric case we have $M_\Phi=M_r$ since the 
equipotential surface $R(\Phi,\theta)$ and the density 
$\rho(r,\theta)$ do not depend upon colatitude $\theta$.

This mass expression sets up a one-to-one relationship between mass 
$M_\Phi$ and equipotential $\Phi$:
\begin{equation}
  m\equiv M_\Phi=M_\Phi(\Phi),
\end{equation}
which permits us to use mass $M_\Phi$ and colatitude $\theta$ as 
the 2D independent variables, instead of the gravitational 
potential $\Phi$ and colatitude $\theta$. Eq.~(\ref{eq:mass1}) is 
the integral form of the mass conservation. Its differential form 
can be obtained by taking its partial derivative with respect to 
the equipotential surface $r=R(\Phi,\theta)$:
\begin{equation}
  \p{m}{r}=\p{M_\Phi}{R} = 4\pi r^2\rho_m, \label{eq:mass2}
\end{equation}
where
\begin{equation}
  \rho_m \equiv \frac{1}{2r^2}\int_0^\pi d\theta R^2(\Phi,\theta)
    \rho(R(\Phi,\theta),\theta)\sin\theta. \label{eq:dm}
\end{equation}
This defines the density on the equipotential surface 
$r=R(\Phi,\theta)$. It should be pointed out that here $r$ is no 
longer a static Eulerian space coordinate, but a co-moving 
Lagrangian variable with an equipotential surface $\Phi=$ constant, 
as it is in the spherically symmetric case. Obviously, $\rho_m=\rho$
when the system is spherically symmetric. We use $r=r(m,\theta)$ to 
denote the functional relationship between $r$ and $m$.

Eq.~(\ref{eq:mass2}) is our 2D mass conservation equation. 
Comparing it with its 1D counterpart,
\begin{equation}
  \p{m}{r} = 4\pi r^2 \rho, \label{eq:mass1d}
\end{equation}
we find that they differ by a correction factor $\rho_m/\rho$:
\begin{equation}
  \p{m}{r} = 4\pi r^2 \rho\left\{\frac{\rho_m}{\rho}\right\}. 
\label{eq:mass2d}
\end{equation}
This factor equals unity in the spherically symmetric case, but 
deviates from unity in the general case.

\subsection{Momentum Conservation}

When both turbulence and magnetic fields are taken into account, the 
momentum conservation of an equilibrium state can be expressed by 
the momentum equation
\ba
&&  \nabla\cdot\left[\left(P+\frac{B^2}{8\pi}+\rho v''_rv''_r\right)\vb{I} 
+ \rho(v''_\theta v''_\theta-v''_rv''_r)\unit{\theta}\unit{\theta} \right.\nonumber \\
 &&\left. + \rho(v''_\phi v''_\phi-v''_rv''_r)\unit{\phi}\unit{\phi}
- \frac{1}{4\pi}\vb{B}\vb{B}\right] 
   = -\rho\nabla\Phi - \nabla\cdot(\rho \vb{v}\vb{v}), \label{eq:mom1}
\ea
where $P$ is the gas pressure, $\vb{B}$ is the magnetic field, 
$\vb{I}$ is the unit tensor with nonzero components 
$\unit{r}\unit{r}$, $\unit{\theta}\unit{\theta}$ and 
$\unit{\phi}\unit{\phi}$, and $v''_i$ is 
the turbulent velocity that is defined by the velocity variance:
\begin{equation}
  v''_i =(\overline{v_i^2}-\overline{v_i}^2)^{1/2},
\end{equation}
where $i=r,\theta,\phi$.
The over-bar denotes a combined horizontal and temporal average, and 
$v_i$ is the total velocity component. See Robinson et al 
(2003) for the details of 3D simulations to derive realistic turbulence 
properties in the solar convection zone, where $v''_\theta=v''_\phi$ is assumed.
The regular motion velocity is denoted by $\vb{v}$, for example, $\vb{v}=\vb{\Omega}\times\vb{r}$ 
for rotation, where $\vb{\Omega}$ is the rotation angular velocity.

For a system with magnetic fields, turbulence and rotation, Eq.~(\ref{eq:mom1}) can be rewritten as follows:
\begin{eqnarray}
  \p{P_T}{r} &=& -\rho \p{\Phi}{r} + \rho(\h{H}_r + \h{T}_r + \h{R}_r), \label{eq:momr}\\
  \frac{1}{r}\p{P_T}{\theta} &=& -\frac{\rho}{r} \p{\Phi}{\theta} 
+ \rho(\h{H}_\theta + \h{T}_\theta + \h{R}_\theta), \label{eq:moma}
\end{eqnarray}
where the isotropic pressure components of the magnetic field 
$\vb{B}$, $P_m=B^2/8\pi$, and the radial pressure component of 
turbulence, $P_t=\rho v''_rv''_r$ have been added to the gas 
pressure, $P$, to define a total isotropic pressure 
$P_T=P+P_m+P_t$, while their anisotropic pressure components are 
denoted by $\h{H}=\frac{1}{4\pi\rho}\nabla\cdot(\vb{B}\vb{B})$ for the 
magnetic field $\vb{B}$, $\h{T}=\rho^{-1}\nabla\cdot[\rho(v''_rv''_r-v''_\theta 
v''_\theta)\unit{\theta}\unit{\theta}+\rho(v''_rv''_r-v''_\phi 
v''_\phi)\unit{\phi}\unit{\phi}]$ for turbulence, 
and $\h{R}=-\rho^{-1}\nabla\cdot(\rho\vb{v}\vb{v})$ for rotation, where 
$\vb{v}=\Omega\times\vb{r}$. Their $r$- and $\theta$-components are:
\begin{mathletters}
\begin{eqnarray}
 4\pi\rho \h{H}_r &=& \frac{1}{r^2}\p{}{r}(r^2B_r^2) 
+\frac{1}{r}\p{}{\theta}(B_rB_\theta)-\frac{1}{r}(B_\theta^2+B_\phi^2), \label{eq:hr} \\
 4\pi\rho \h{H}_\theta &=&\frac{1}{r}\p{B_\theta^2}{\theta} 
  +\frac{1}{r}\p{}{r}(rB_rB_\theta) -\frac{B_\phi^2}{r}\cot\theta, \label{eq:ha} \\
  \h{T}_r &=&\frac{2}{r}(v''_\theta v''_\theta-v''_r v''_r), \label{eq:tr} \\
  \h{T}_\theta &=&-\rho^{-1}\frac{1}{r}\p{}{\theta}[\rho(v''_\theta v''_\theta-v''_r v''_r)], \label{eq:ta} \\
  \h{R}_r &=& \Omega^2 r \sin^2\theta, \label{eq:rr} \\
  \h{R}_\theta &=& \Omega^2 r\sin\theta\cos\theta. \label{eq:ra} 
\end{eqnarray}
\end{mathletters} 

\subsection{Poisson Equation}

The Poisson equation in the spherical 
coordinate system with the specified symmetry requirement 
can be written down as follows :
\begin{equation}
\frac{1}{r^2}\p{}{r}\left(r^2 \p{\Phi}{r}\right) 
+\frac{1}{r^2\sin\theta}\p{}{\theta}\left(\sin\theta 
\p{\Phi}{\theta}\right) =4\pi G\rho. \label{eq:poisson}
\end{equation}
Solving this equation for the gravitational potential $\Phi$
is not sufficiently accurate for our purposes, especially in the core of 
stars. Solving it for the radial gravitational acceleration 
$g=\partial\Phi/\partial r$ is equally not good for the same reason. 
Many tries show that the following treatment is 
sufficiently accurate for our high-precision requirement.

First of all, we calculate the colatitudinal gravitational 
acceleration $\h{G}\equiv(1/r)\partial\Phi/\partial\theta$ by using 
the hydrostatic equilibrium equation in the colatitudinal 
direction (Eq.~\ref{eq:moma}) in terms of $\partial 
P_T/\partial\theta$, $\h{H}_\theta$, and $\h{T}_\theta$:
\begin{equation}
  \h{G} = \h{H}_\theta + \h{T}_\theta + \h{R}_\theta
    -\frac{1}{r\rho}\p{P_T}{\theta}. \label{eq:gg1}
\end{equation}
This way, Eq.~(\ref{eq:moma}) is satisfied automatically. We then 
decompose $g$ into two parts,
\begin{equation}
  g=\frac{Gm}{r^2} + \delta g. \label{eq:g}
\end{equation}
The first part is the spherically-symmetric radial component of the 
gravitational acceleration, and the second part is the deviation of 
the radial gravitational acceleration from its 
spherically-symmetric counterpart. Substituting Eq.~(\ref{eq:g}) 
into Eq.~(\ref{eq:poisson}), we obtain
\begin{equation}
  \p{\delta g}{r} = 4\pi G(\rho-\rho_m) - \frac{2}{r}\delta g 
     -\frac{\h{G}\cot\theta}{r}-\frac{1}{r}\p{\h{G}}{\theta}.     \label{eq:dg}
\end{equation}
Therefore, we solve the Poisson equation for $\delta g$ instead of 
$\Phi$ or $g$. Here we have used the notations $r'=\ln r$, and $\rho'=\ln\rho$.
The hydrostatic equilibrium equation in the radial 
direction thus becomes:
\begin{equation}
  \p{P_T}{r} =-\rho\left(\frac{Gm}{r^2} +\delta g - \h{H}_r -\h{T}_r - \h{R}_r\right). \label{eq:momr1}
\end{equation}

\subsection{Energy Conservation}

The energy conservation equation is
\begin{equation}
\frac{1}{r^2}\p{}{r}(r^2 F_r) +\frac{1}{r\sin\theta}\p{}{\theta}
  (\sin\theta F_\theta) =\rho\left(\epsilon-T\od{S_T}{t}\right), 
  \label{eq:energy2d}
\end{equation}
where $\vb{F}=\vb{F}_{\s{rad}}+\vb{F}_{\s{conv}}$ is the energy 
flux vector, including both the radiative flux $\vb{F}_{\s{rad}}$ 
and the convective flux $\vb{F}_{\s{conv}}$, and  $\epsilon$ is the 
rate of nuclear energy generation, and $S_T$ is the total specific 
entropy, including the contributions from magnetic fields and 
turbulence. We use the diffusion approximation for radiative flux, 
and the mixing length theory for convective flux:
\begin{eqnarray}
\vb{F}_{\s{rad}} &=& -\frac{4acT^3}{3\kappa\rho}\nabla T, \label{eq:fluxr} \\
\vb{F}_{\s{conv}} &=& -\frac{1}{2}\frac{\rho Tl_m v_{\s{conv}}}{1+v_{\s{conv}}/v_0}\nabla S_T, \label{eq:fluxa}
\end{eqnarray}
where $v_{\s{conv}}$ is the convection velocity, $l_m$ is the 
mixing length, $v_0$ is a typical velocity determined by choice of 
radiative loss mechanism of a convective eddy. The symbol $a$ 
represents the radiation constant, $c$ the speed of light, $\kappa$ the 
mass opacity coefficient. The 2D energy conservation equation shows 
that energy can not only penetrate a region via the radial gradient 
of the radial component of the energy flux, but also goes around it 
via the transverse gradient of the transverse component of the 
energy flux. In contrast, the 1D energy conservation equation
\begin{equation}
 \frac{1}{r^2}\od{}{r}(r^2 F_r) = \rho\left(\epsilon-T\od{S_T}{t}\right) \label{eq:energy1d}
\end{equation}
rules out the transverse transport of energy.
 
\subsection{Energy Transport}

Eqs.~(\ref{eq:energy2d}-\ref{eq:energy1d}) show that we have to 
calculate temperature and entropy gradients. We thus need the first 
law of thermodynamics in the presence of magnetic fields and 
turbulence. We have redefined the mechanical variable $P_T$ by 
adding all isotropic pressure components together. We need magnetic 
and turbulent variables to take into account magnetic and turbulent 
degrees of freedom.

\subsubsection{Magnetic and turbulent variables}

We use $\vb{B}$ to define three stellar magnetic parameters, in 
addition to the conventional stellar parameters such as pressure, 
temperature, radius and luminosity. The first magnetic parameter is 
the magnetic kinetic energy per unit mass, $\chi_m$,
\begin{equation}
   \chi_m = B^2/(8\pi\rho).
\end{equation}
The second is the heat index due to the magnetic field, or the 
ratio of the magnetic pressure in the radial direction to the 
magnetic energy density, $\gamma_m-1$,
\begin{equation}
  \gamma_m = 1 + (B_\theta^2+B_\phi^2)/B^2.
\end{equation}
The third one is the ratio of the magnetic pressure in the 
colatitude direction to the magnetic energy density, $\vartheta_m-1$,
\begin{equation}
 \vartheta_m=1+(B_\phi^2+B_r^2)/B^2.
\end{equation}
We can use these three magnetic parameters to express 
three components of a magnetic field as follows:
\begin{mathletters}
\begin{eqnarray}
   B_r &=& [8\pi(2-\gamma_m) \chi_m\rho]^{1/2}, \label{eq:br0} \\
   B_\theta &=& [8\pi(2-\vartheta_m)\chi_m\rho]^{1/2}, \label{eq:bt0} \\
   B_\phi   &=& [8\pi(\gamma_m+\vartheta_m-3)\chi_m\rho]^{1/2}. \label{eq:bp0}
\end{eqnarray}
\end{mathletters} 
However, since $v''_\theta=v''_\phi$ is assumed, we have only two 
turbulent degrees of freedom and we thus need two turbulent 
variables, namely, the turbulent kinetic energy per unit mass, 
$\chi_t$, and the effective ratio of specific heats due to 
turbulence, $\gamma_t$:
\begin{eqnarray} 
  \chi_t = \frac{1}{2} (v'')^2, && 
  \gamma_t = 1 + 2(v''_r/v'')^2.
\end{eqnarray}
We can use them to express three turbulent velocity components:
\begin{mathletters}
\begin{eqnarray}
   v''_r &=& [(\gamma_t-1)\chi_t]^{1/2}, \label{eq:vr0} \\
   v''_\theta &=& v''_\phi=\left[\frac{1}{2}(3-\gamma_t)\chi_t\right]^{1/2}. \label{eq:vt0}
\end{eqnarray}
\end{mathletters} 

\subsubsection{Equation of state}

Using the magnetic and turbulent variables defined above, we can 
rewrite the total pressure as follows:
\begin{equation}
  P_T = P(\rho,T) + \rho\chi_m + \rho(\gamma_t-1)\chi_t.
\end{equation}
Solving this equation for $\rho$, we obtain the equation of state 
in the presence of magnetic fields and turbulence:
\begin{equation}
  \rho=\rho(P_T,T,\chi_m,\chi_t,\gamma_t). \label{eq:eos}
\end{equation}
To highlight magnetic and turbulence effects we adopt a given  
chemical composition. This shows that the independent 
thermodynamical variables are $P_T$, $T$, $\chi_m$, $\chi_t$, and 
$\gamma_t$. Using them,we can write the differential form of the equation of 
state as follows:
\begin{equation}
  d\rho/\rho=\alpha dP_T/P_T -\delta dT/T -\nu_m d\chi_m/\chi_m 
-\nu_t d\chi_t/\chi_t-\mu_t d\gamma_t/\gamma_t,
\end{equation}
where
\begin{mathletters}
\begin{eqnarray}
&&  \alpha \equiv (\partial\ln\rho/\partial\ln P_T)_{T,\chi_m,\chi_t,\gamma_t}, \hspace{5mm} \delta \equiv -(\partial\ln\rho/\partial\ln T)_{P_T,\chi_m,\chi_t,\gamma_t}, \label{def:alpha} \\
&&  \nu_m \equiv -(\partial\ln\rho/\partial\ln \chi_m)_{P_T,T,\chi_t,\gamma_t}, \hspace{5mm} \nu_t \equiv -(\partial\ln\rho/\partial\ln\chi_t)_{P_T,T,\chi_m,\gamma_t}\\
&& \mu_t \equiv -(\partial\ln\rho/\partial\ln\gamma_t)_{P_T,T,\chi_m,\chi_t}. 
\label{def:beta}
\end{eqnarray}
\end{mathletters}

When a $\theta$-dependent magnetic field is applied, Eq.~(\ref
{eq:eos}) demonstrates that the mass distribution will adjust to 
generate asphericity. This is the most straightforward 2D effect.

\subsubsection{The first law of thermodynamics in the presence 
of magnetic fields and turbulence}

The first law of thermodynamics is the energy transfer and 
conservation law in a thermodynamic system. In the presence of 
magnetic fields and turbulence, the conservation law should be 
modified as follows:
\begin{equation}
  TdS_T = dU + PdV -d\chi_m - d\chi_t, \label{eq:law1}
\end{equation}
which states that both magnetic and turbulent energy are generated 
at the expense of internal energy of the system $U$. Here $V=1/\rho$
is the specific volume. Combining Eqs.~(\ref{eq:eos}) and 
(\ref{eq:law1}) (see Lydon \&\ Sofia 1995 for the detail), we obtain
\begin{equation}
  TdS_T = C_p dT -\left(\frac{\delta}{\rho}\right)dP_T 
+\left(\frac{P_T\delta\nu_m}{\alpha\rho\chi_m}-1\right)d\chi_m
   +\left(\frac{P_T\delta\nu_t}{\alpha\rho\chi_t}-1\right)d\chi_t
+\frac{P_T\delta\mu_t}{\alpha\rho\gamma_t}d\gamma_t,   \label{eq:law1a}
\end{equation}
from which we obtain
\begin{eqnarray}
  \nabla S_T &=& (C_p/T)\nabla T - (C_p\nabla'_{\s{ad}}/P_T)\nabla P_T, \label{eq:law1b} \\ 
  \od{S_T}{t} &=& (C_p/T)\od{T}{t} - (C_p\nabla'_{\s{ad}}/P_T)\od{P_T}{t}.\label{eq:law1c}
\end{eqnarray}
We have defined the modified adiabatic gradient
\begin{equation}
  \nabla'_{\s{ad}}=\nabla_{\s{ad}}\left[1 - \left(\frac{\nu_m}{\alpha}
-\frac{\chi_m}{C_pT}\right)\nabla_m
  -\left(\frac{\nu_t}{\alpha}-\frac{\chi_t}{C_pT}\right)\nabla_t 
- \frac{\mu}{\alpha}\nabla_\gamma\right], \label{eq:adiabatic}
\end{equation}
where $C_p$ is the specific heat per unit mass at constant total 
pressure, constant magnetic energy per unit mass, constant 
turbulent kinetic energy per unit mass, and constant turbulent 
specific heat ratio, and
\[
 \nabla_{\s{ad}}=P_T\delta/(\rho C_pT), \hspace{3mm} \nabla_m=\p{\ln \chi_m}{\ln P_T}, 
 \hspace{3mm} \nabla_t=\p{\ln\chi_t}{\ln P_T}, 
\hspace{3mm} \nabla_\gamma=\p{\ln\gamma_t}{\ln P_T}.
\]
The physical meaning of Eq.~(\ref{eq:adiabatic}) is that magnetic 
fields and turbulence provide additional channels for energy 
transport.

\subsubsection{Energy flux vector}

Using Eqs.~(\ref{eq:fluxr}-\ref{eq:fluxa}) and (\ref{eq:law1b}) the 
energy flux vector $\vb{F}$ can be expressed by the temperature 
gradient $\nabla T$ and pressure gradient $\nabla P_T$ as follows:
\begin{eqnarray}
\vb{F} &=& -\left(\frac{4ac T^3}{3\kappa\rho} 
     + \frac{1}{2} \frac{\rho C_P l_m v_{\s{conv}}}{1+v_{\s{conv}}/v_0}\right)  \nabla T \nonumber\\
  &&  + \frac{1}{2} \frac{\rho C_P T \nabla'_{\s{ad}} l_m v_{\s{conv}}}{1+v_{\s{conv}}/v_0}
\frac{1}{P_T}\nabla P_T.  \label{eq:flux2d}
\end{eqnarray}
Its $r$-component determines the radial temperature gradient, 
its $\theta$-component results in 2D effect.

\section{EQUIPOTENTIAL SURFACE}\label{sct:s3}

Solar magnetic fields are weak in the sense that the resultant magnetic 
pressure is much smaller than the gas pressure. The usual central difference 
scheme alone may not discern the required 2D effects. We should therefore use
certain physical guidelines to improve the precision of the numerical solutions
for the 2D stellar structure equations. The key concept introduced for the 2D stellar 
structure in this series is the equipotential surface.
In this section we show how to determine it.

\subsection{Exact 2D Stellar Structure Equations}

The exact 2D stellar structure equations, i.e., Eqs.~(\ref
{eq:mass2d}), (\ref{eq:gg1}), (\ref{eq:dg}), (\ref{eq:momr1}), (\ref
{eq:energy2d}), and  the energy transport equation, can be 
rewritten as follows after coordinate transformation from 
$(r,\theta)$ to $(m,\theta)$:
\begin{mathletters}
\ba
\p{r'}{s}  &=& \frac{m}{4\pi r^3\rho}\frac{\rho}{\rho_m}, \label{eq:st1} \\
\p{P'}{s}&=& -\frac{m}{4\pi r^2 P_T}\frac{\rho}{\rho_m}\left(\frac{Gm}{r^2} 
    +U-\h{H}_r-\h{T}_r-\h{R}_r\right), \label{eq:st2} \\
\p{T'}{s} &=& \p{P'}{s} \left\{ \begin{array}{ll}
         \nabla_{\s{rad}} & \mbox{ radiative} \\
         \nabla_{\s{c}} & \mbox{ convective} \\
    \end{array} \right. \label{eq:st3} \\
\p{L}{s} &=& \frac{1}{L_\sun}m \left(\epsilon-T\od{S_T}{t}\right)
  \frac{\rho}{\rho_m}
  -\frac{1}{L_\sun}\frac{mF_\theta\cot\theta}{r\rho_m} 
  -\frac{1}{L_\sun}\frac{m}{r\rho_m}\p{F_\theta}{\theta}, \label{eq:st4} \\
\p{U}{s} &=& \frac{Gm}{r^2}\left(\frac{\rho}{\rho_m}-1\right)
   -\frac{m}{4\pi r^3\rho_m}\left(2 U +\h{G}\cot\theta +\p{\h{G}}{\theta}\right). \label{eq:st5}
\ea
\end{mathletters}
Here $P'=\ln P_T$, $T'=\ln T$, $r'=\ln r$, $L=4\pi r^2 F_r/L_\sun$, and 
$U=\delta g$. The other symbols used above are defined as follows:
\begin{mathletters}
\ba
F_\theta &=&  \left\{-\left[\frac{4ac T^4}{3\kappa\rho}
  + \frac{1}{2} \frac{\rho C_PT l_m v_{\s{conv}}}{1+v_{\s{conv}}/v_0}\right]  
  \frac{\nabla}{r} + \frac{1}{2} \frac{\rho C_P T l_m v_{\s{conv}}}{1+v_{\s{conv}}/v_0}
  \frac{\nabla'_{\s{ad}}}{r}\right\}\p{P'}{\theta}, \label{eq:q3} \\
\h{G} &=&  \h{H}_\theta+\h{T}_\theta + \h{R}_\theta
   -\frac{P_T}{r\rho}\p{P'}{\theta}. \label{eq:gg}
\ea
\end{mathletters}

These equations show that in addition to the dependent variables, 
pressure $P_T$, temperature $T$, radius $r$, and luminosity $L$, we 
have two more dependent variables, the radial and colatitudinal 
gravitational acceleration perturbations $\delta g$ and $\h{G}$. 
However, we need to solve only five partial differential equations 
(Eqs.~\ref{eq:st1}-\ref{eq:st5}) because the colatitudinal 
gravitational acceleration $\h{G}$ can be calculated by using 
$\pa{P_T}{\theta}$, $\pa{r'}{\theta}$, $\h{H}_\theta$, 
$\h{T}_\theta$, and $\h{R}_\theta$.

We use $\delta g=0$ at $m=0$ as the central boundary condition for 
the fifth equation because $\delta g$ is a perturbation in nature.
This is equivalent to assume that the radial gravitational acceleration 
be equal to its spherically symmetric counterpart at the center.

\subsection{Equipotential Surface Profile}

We know that $r$ is the radial coordinate of an equipotential surface. Its
dependence on the colatitudinal coordinate $\theta$, i.e.,
$r=r(\Phi,\theta)=r(m,\theta)$ defines an equipotential surface on which the
potential equals $\Phi$. We redefine $r$ by $r_e=r_e(m)$ and $x=x(m,\theta)$:
$r=r_e(m)x(m,\theta)$, where $r_e$ is the equatorial radius. Since we interpret $r_e$ 
as the equatorial radius, $x$ should always be normalized so that we obtain $x=1$ at 
the equator where $\theta=\pi/2$. The equipotential surface is thus expressed 
by $x=x(m,\theta)$, which is a function of mass $m=M_\Phi(\Phi)$ and colatitude $\theta$.
 
In order to find out the equipotential surface $x$, we use the fact that pressure
is $\theta$-independent on it. Otherwise, the hydrostatic equilibrium is not
reached thereon. This indicates that $\p{P_T}{s}$ should be $\theta$-independent
thereon as well. The following equation is $\theta$-independent and holds well for both
spherically-symmetric and aspherical cases:
\be
   \p{P'}{s} = -\frac{Gm^2}{4\pi r_e^4 P_T}. \label{eq:p1}
\ee
Comparing it with Eq.~(\ref{eq:st2}), we obtain
\be
   x^4 = \frac{\rho}{\rho_m}\left[1
+\frac{r_e^2x^2}{Gm}\left(U-\h{H}_r
    -\h{T}_r-\h{R}_r\right)\right]. \label{eq:x1}
\ee
We can use the iteration method to obtain $x$ starting from $x=1$ if we 
know $r_e$, $U$, $\h{H}_r$, $\h{T}_r$, and $\h{R}_r$. Eq.~(\ref{eq:dm}) that is
used to determine $\rho/\rho_m$ now becomes
\be
  \rho_m = \frac{1}{x^2}\int_0^{\pi/2} \rho x^2 \sin\theta d\theta. \label{eq:rhom1}
\ee

\subsubsection{Mass conservation for $r_e$}

To calculate $r_e$, we rewrite the mass conservation equation as follows:
\be
  \p{r}{m}=1/Q,
\ee
where
\be
   Q \equiv 4\pi r^2\rho_m
\ee
is $\theta$-independent. As a result, we know that $\p{r}{m}$ is 
$\theta$-independent. Therefore, we can choose $r$ at any specific colatitude on
the equipotential surface. We can, of course, choose $r=r_e(m)$ to obtain
\be
  \p{r_e'}{s} = \frac{m}{Qr_e}, \label{eq:r1}
\ee
where $r'_e \equiv \ln r_e$.

Eq.~(\ref{eq:r1}) becomes
\be
  m = \frac{1}{3} Qr_e
\ee
at the center. This is one of the central boundary conditions.

\subsubsection{Poisson equation for $U$}

The radial gravitational acceleration perturbation $U=\delta g$ can be decomposed 
into five components $U=U_D + U_P + U_H + U_T + U_R$ according to their 
physical origins specified by the subscripts, where subscript D stands for
the density variation, P for the pressure variation, H for magnetic fields, 
T for turbulence, and R for rotation. To see this, we decompose the colatitudinal 
gravitational acceleration component into four components according to their
physical ingredients $\h{G} = \h{G}_P + \h{G}_H + \h{G}_T + \h{G}_R$. Their
definitions are
\begin{mathletters}
\ba
  \h{G}_P &=& -\frac{GmQ}{4\pi r_e^4\rho}\pa{x'}{\theta}, \\
  \h{G}_H &=& \h{H}_\theta, \\
  \h{G}_T &=& \h{T}_\theta, \\
  \h{G}_R &=& \h{R}_\theta,
\ea
\end{mathletters}
where we have utilized the equipotential surface condition $\pa{P'}{\theta}=0$ 
and defined $x'\equiv\ln x$. Since the Poisson equation is linear, we can write 
it down for each component as follows:
\be
  \p{U_i}{r} = -\frac{2U_i}{r} + S_i, \label{eq:pui}
\ee
where i = D, P, H, T, and R. The source terms $S_i$ are expressed by the following functions:
\begin{mathletters}
\ba
  S_D &=& 4\pi G(\rho-\rho_m), \label{eq:u1} \\
  S_i &=& - \frac{\h{G}_i\cot\theta}{r} -\frac{1}{r}\p{\h{G}_i}{\theta}, \label{eq:u2}
\ea
\end{mathletters}
where i = P, H, T, and R. Eq.~(\ref{eq:pui}) has a specific solution:
\be
  U_i = \frac{1}{x^2}\int_0^r x^2S_i dr. \label{eq:ui}
\ee

\subsection{Uniform Rotation Rate}

\subsubsection{Uniform rotation equipotential surface}

We want to use this special case to show how to obtain the equipotential 
surface $x=x(m,\theta)$.

For rotation at the angular velocity $\Omega \hat{z}$, we can use 
Eq.~(\ref{eq:ui}) to calculate the radial gravitational acceleration perturbation
$U_R$. The result is
\Ba
  \h{G}_R &=& \h{R}_\theta = \frac{1}{2}\Omega^2r\sin2\theta, \\
  S_R &=& - \frac{3}{2}\Omega^2(\cos2\theta+\frac{1}{3}), \\
  U_R &=& -\frac{3}{2}\Omega^2 r(\cos2\theta+\frac{1}{3}).
\Ea
Here we assume that $\Omega=\Omega(r)$ does not depend upon $\theta$.
Eq.~(\ref{eq:x1}) shows that we need
\[
U_R-\h{R}_r=-\Omega^2 r(\cos2\theta+1).
\]
As the first approximation, we assume $\rho/\rho_m=1$, $x=1$, and $U_P=U_D=0$ 
in Eq.~(\ref{eq:x1}). For a slow rotation in the sense that the centrifugal 
acceleration $\Omega^2r_e$ is much smaller than the corresponding gravitational
acceleration $Gm/r_e^2$, we obtain
\be
   x^{(0)} = 1 - \frac{1}{4}a_0 (\cos2\theta+1), \label{eq:l0}
\ee
where
\[
a_0=\frac{\Omega^2r^3}{Gm}. 
\]

We can further improve the result by taking into account 
$\varpi\equiv\rho/\rho_m$, $\Lambda\equiv\pa{x'}{\theta}$, $U_P$, and $U_D$ in
Eq.~(\ref{eq:x1}):
\Ba
  \varpi^{(0)} &=& 1-\frac{1}{2}a_0(\cos2\theta+\frac{1}{3}), \\
  \Lambda^{(0)} &=&  \frac{1}{2}a_0\sin2\theta, \\
  \h{G}_P^{(0)} &=& - \frac{1}{2}\frac{Gm a_0}{r^2}\sin2\theta, \\
  S_P^{(0)} &=& \frac{1}{2}(\cos2\theta+\frac{1}{3})\frac{3Gm a_0}{r^3}, \\
  S_D^{(0)} &=& -\frac{1}{2}(\cos2\theta+\frac{1}{3})4\pi Ga_0\rho, \\
  U_P^{(0)} &=& \frac{1}{2}(\cos2\theta+\frac{1}{3})b_P^{(0)} , \\
  U_D^{(0)} &=& -\frac{1}{2}(\cos2\theta+\frac{1}{3})b_D^{(0)},
\Ea
where 
\[
  b_P^{(0)}=3\int_0^r\frac{Gm a_0}{r^3}dr, \hspace{3mm} b_D^{(0)} =  4\pi G\int_0^r a_0\rho dr.
\]
The corrected equipotential surface function is
\be
  x^{(1)} = 1 +\frac{1}{12}[a_0+\frac{r^2}{Gm}(b_D^{(0)}-b_P^{(0)})]-\frac{1}{4}a_1(\cos2\theta+1), 
\ee
where
\[
  a_1=a_0 + a_0',\hspace{3mm} a_0'= \frac{1}{2}[a_0
    +\frac{r^2}{Gm}(b_D^{(0)}-b_P^{(0)})].
\]
According to the definition of $x$, it should equal unity at the equator. This requirement
fixes the expression of $x$ as follows:
\be
  x^{(1)} = 1 -\frac{1}{4}a_1(\cos2\theta+1). \label{eq:l1}
\ee
From now on, we shall show this form only, which will be referred to as 
the {\it normalized form}.

Using Eq.~(\ref{eq:l1}) or its non-normalized form we can improve $\varpi$, 
$\Lambda$, $U_P$, and $U_D$:
\Ba
  \varpi^{(1)} &=& 1-\frac{1}{2}a_1(\cos2\theta+\frac{1}{3}), \\
  \Lambda^{(1)} &=&  \frac{1}{2}a_1\sin2\theta, \\
  \h{G}_P^{(1)} &=& - \frac{1}{2}\frac{Gm a_1}{r^2}\sin2\theta, \\
  S_P^{(1)} &=& \frac{1}{2}(\cos2\theta+\frac{1}{3})\frac{3Gm a_1}{r^3}, \\
  S_D^{(1)} &=& -\frac{1}{2}(\cos2\theta+\frac{1}{3})4\pi Ga_1\rho, \\
  U_P^{(1)} &=& \frac{1}{2}(\cos2\theta+\frac{1}{3})b_P^{(1)}, \\
  U_D^{(1)} &=& -\frac{1}{2}(\cos2\theta+\frac{1}{3})b_D^{(1)},
\Ea
where
\[
 b_P^{(1)}=3\int_0^r \frac{Gma_1}{r^3}dr, 
  \hspace{3mm} b_D^{(1)}=4\pi G\int_0^r a_1\rho dr.
\]
The more accurate equipotential surface is thus expressed by
\be
  x^{(2)} = 1 -\frac{1}{4}a_2(\cos2\theta+1), \label{eq:l2}
\ee
where
\[
  a_2 = a_0 + a_1', \hspace{3mm} a_1'=\frac{1}{2}[a_1
   +\frac{r^2}{Gm}(b_D^{(1)}-b_P^{(1)})].
\]

To keep iterating, we find the following recurrence relation 
for i = 1, 2, 3, $\cdots$:
\be
  x^{(i)} = 1 -\frac{1}{4}a_i(\cos2\theta+1), \label{eq:ri}
\ee
where
\begin{mathletters}
\ba
  a_i &=& a_0+a'_{i-1}, \\
  a_i' &=& \frac{1}{2}[a_i+\frac{r^2}{Gm}(b_D^{(i)}-b_P^{(i)})],\\
  b_P^{(i)} &=& 3\int_0^r \frac{Gma_i}{r^3}dr, \\
  b_D^{(i)} &=& 4\pi G\int_0^r a_i\rho dr.
\ea
\end{mathletters}
Using the equipotential surface profile, Eq.~(\ref{eq:ri}), we can calculate the
following quantities:
\begin{mathletters}
\ba
  Q^{(i)} &=& 4\pi r_e^2\rho(1-\frac{1}{3}a_i), \\
  \varpi^{(i)} &=& 1-\frac{1}{2}a_i(\cos2\theta+\frac{1}{3}), \\
  \Lambda^{(i)} &=&  \frac{1}{2}a_i\sin2\theta, \\
  \h{G}_P^{(i)} &=& - \frac{1}{2}\frac{Gm a_i}{r^2}\sin2\theta, \\
  U_P^{(i)} &=& \frac{1}{2}(\cos2\theta+\frac{1}{3})b_P^{(i)}, \\
  U_D^{(i)} &=& -\frac{1}{2}(\cos2\theta+\frac{1}{3})b_D^{(i)}.
\ea
\end{mathletters}
The gravitational acceleration perturbations due to rotation are
\begin{mathletters}
\ba
  \h{G}^{(i)} &=& \frac{1}{2}(\Omega^2r - \frac{Gm a_i}{r^2})\sin2\theta, \\
  U^{(i)} &=& -\frac{1}{2}[3\Omega^2r+(b_D^{(i)}-b_P^{(i)})](\cos^2\theta+\frac{1}{3}).
\ea
\end{mathletters}
With inclusion of the rotation effects, the gravitational acceleration vector 
can be expressed as follows:
\begin{mathletters}
\ba
  g_r^{(i)} &=& \frac{Gm}{r^2} -\frac{1}{2}[3\Omega^2r+(b_D^{(i)}-b_P^{(i)})]
   (\cos2\theta+\frac{1}{3}), \label{eq:rgr}\\
  g_\theta^{(i)} &=& \frac{1}{2}\left(\Omega^2r 
- \frac{Gm a_i}{r^2}\right)\sin2\theta. \label{eq:rga}
\ea
\end{mathletters}
We know $r=r_ex^{(i)}$. Since $b_P^{(i)}$ and $b_D^{(i)}$ are integrals over $r$ from $0$ to $r$, 
we know that the gravitational acceleration perturbations $U^{(i)}$ and $\h{G}^{(i)}$ do not 
vanish outside the star.

\subsubsection{Uniform rotation-like magnetic equipotential surface}\label{sct:um}

Rotation has a global velocity field $\vb{v}=(0,0,\Omega r\sin\theta)$. We can choose a
toroidal magnetic field $\vb{B}=(0,0,(4\pi\rho)^{1/2}\Omega r\sin\theta)$ to mimic
rotation at the rate $\Omega$. We use this magnetic configuration to show the
calculation method for the magnetic equipotential surface and the difference between
rotation and magnetic effects.

The first step is to calculate two components of $\h{H}$: $\h{H}_r$ and $\h{H}_\theta$.
They are
\Ba
  \h{H}_r &=& - \Omega^2 r\sin^2\theta, \\
  \h{H}_\theta &=& -\Omega^2 r\sin\theta\cos\theta.
\Ea
Comparing them with the corresponding $\h{R}_r$ and $\h{R}_\theta$, we can see that 
their signs are opposite.

We also need the plasma $\beta$ parameter. Its definition is the ratio of the total pressure $P_T$ over the magnetic pressure 
$P_m=\frac{1}{2}\rho\Omega^2 r^2\sin^2\theta$. Using 
$\beta_0=2P_T/\rho_0\Omega^2 r^2$, we have $\beta=\beta_0/\sin^2\theta$. Magnetic
pressure causes a density change. The density ($\rho/\rho_0$) with/without the magnetic
field is related to each other by the formula $\rho=\rho_0/(1+1/\beta)$, or
$\rho=\rho_0/(1+c_2\sin^2\theta)$, where we have used $c_2=1/\beta_0$ to replace
$\beta_0$. We know $c_2=\rho_0r^2\Omega^2/2P_T$.

The next step is to use $\h{G}_H=\h{H}_\theta$ to obtain the source term $S_H$:
\[
  S_H = \frac{3}{2}\Omega^2(\cos2\theta+\frac{1}{3}).
\]
Substituting it into Eq.~(\ref{eq:ui}), we obtain
\[
  U_H = \frac{3}{2}\Omega^2 r(\cos2\theta+\frac{1}{3}).
\]
Using $U_H$ and $\h{H}_r$ in Eq.~(\ref{eq:x1}), we obtain the first 
approximation to the magnetic equipotential surface
\be
  x^{(0)} = 1 + \frac{1}{4} a_0(\cos2\theta+1). \label{eq:h0}
\ee
Comparing Eqs.~(\ref{eq:l0}) and (\ref{eq:h0}), we can see that the oblateness 
$\epsilon=(r_e-r_p)/r_e=\pm a/2$ is positive for rotation, but negative for 
magnetic fields, where $r_p$ is the polar radius.

The following steps differ from the rotation case since the magnetic effect on density, 
which comes from the integral $\rho_m$, cuts in. The density correction to the 
equipotential surface can be expressed by $c_2$ in the recurrence relation
\be
  x^{(i)} = 1+\frac{1}{4}a_i(\cos2\theta+1), \label{eq:hi}
\ee
where
\ba
  a_i &=& a_0+\frac{1}{2}c_2+a_{i-1}', \\
  a_i' &=& \frac{1}{2}\left[a_i +\frac{r^2}{Gm}(b_D^{(i)}-b_P^{(i)})\right], \\
  b_p^{(i)} &=& 3\int_0^r\frac{Gma_i}{r^3}dr, \\
  b_D^{(i)} &=& 4\pi G\int_0^r (a_i+c_2)\rho_0 dr.
\ea

Using the equipotential surface profile, Eq.~(\ref{eq:hi}), we can calculate the
following quantities:
\begin{mathletters}
\ba
  Q^{(i)} &=& 4\pi r_e^2\rho_0(1+\frac{1}{3}a_i-\frac{2}{3}c_2), \\
  \varpi^{(i)} &=& 1 + \frac{1}{2}(a_i+c_2)(\cos2\theta+\frac{1}{3}), \\
  \Lambda^{(i)} &=&  -\frac{1}{2}a_i\sin2\theta, \\
  \h{G}_P^{(i)} &=&  \frac{1}{2}\frac{Gm a_i}{r^2}\sin2\theta, \\
  U_P^{(i)} &=& -\frac{1}{2}(\cos2\theta+\frac{1}{3})b_P^{(i)}, \\
  U_D^{(i)} &=& \frac{1}{2}(\cos2\theta+\frac{1}{3})b_D^{(i)}.
\ea
\end{mathletters}
Since the magnetic effect on density has been totally absorbed into $c_2$, the integrant in
the integral $b_D^{(i)}$ involves $\rho_0$, instead of $\rho=\rho_0/(1+c_2\sin^2\theta)$, which is
the same as above. The gravitational acceleration perturbations due to a rotation-like magnetic
field are
\begin{mathletters}
\ba
  \h{G}^{(i)} &=& -\frac{1}{2}\left(\Omega^2r-\frac{Gm a_i}{r^2}\right)\sin2\theta, \\
  U^{(i)} &=& \frac{1}{2}(3\Omega^2r+b_D^{(i)}-b_P^{(i)})(\cos2\theta+\frac{1}{3}).
\ea
\end{mathletters}
Including the rotation-like magnetic effects, we obtain the expression for the gravitational 
acceleration vector:
\begin{mathletters}
\ba
  g_r^{(i)} &=& \frac{Gm}{r^2} + \frac{1}{2}(3\Omega^2r+b_D^{(i)}-b_P^{(i)})
(\cos2\theta+\frac{1}{3}), \label{eq:hgr} \\
  g_\theta^{(i)} &=& -\frac{1}{2}\left(\Omega^2r-\frac{Gm a_i}{r^2}\right)
\sin2\theta. \label{eq:hga}
\ea
\end{mathletters}

\subsubsection{Uniform rotation-like turbulent equipotential surface}

Solar turbulent data are given by the three-dimensional (3D) numerical simulations 
within a small volume that contains the super-adiabatic layer (SAL) of the Sun. 
The turbulent pressure $P_t=\frac{1}{2}\rho v_r''v_r''$ peaks at the peak of SAL.
The peak value is about 17\% (Robinson et al 2003; Stein \& Nordlund 1998).
Since the simulations are restricted to a small range of the colatitudinal coordinate
and all the turbulent velocity components are the averaged velocity variance over the
colatitudinal coordinate, the $\theta$-dependence of the turbulent velocity is unknown.
Turbulent velocity may have two components, one is $\theta$-independent, and the other
is $\theta$-dependent. The latter must be much smaller than the former.

The $\theta$-independent component has nothing to do with the equipotential surface,
but the $\theta$-dependent component affects the equipotential surface.
In order to address the difference among rotation, magnetic, and turbulent effects, we 
assume that the $\theta$-dependent component of $v_r''v_r''$ equals 
$\frac{1}{2}\Omega^2 r^2 \sin^2\theta$, and that of $v_\theta''v_\theta''$ equals 
zero or $\Omega^2 r^2 \sin^2\theta$. As a result, we have
\begin{mathletters}
\ba
  \h{T}_r &=& \mp\Omega^2 r \sin^2\theta, \\
  \h{T}_\theta &=& \pm\Omega^2 r \sin\theta\cos\theta.
\ea
\end{mathletters}
It is interesting to note that the signs of both $\h{R}_r$ and $\h{R}_\theta$ are 
the same ("+"), those of both $\h{H}_r$ and $\h{H}_\theta$ are the same ("-"), but
those of $\h{T}_r$ and $\h{T}_\theta$ are opposite to each other ("$\mp$" vs "$\pm$").
We have shown above that the sign determines the sign of the oblateness of the
equipotential surface. We thus anticipate something new for turbulence.
Following the same procedure as obtaining Eq.~(\ref{eq:hi}), we obtain 
\be
  x^{(0)} = 1 \pm \frac{1}{4}(2a_0)(\cos2\theta+1). \label{eq:t0}
\ee
The new outcome is that the coefficient doubles, here $a_0=\Omega^2r^3/Gm$ as above.
The recurrence relation thus becomes
\be
  x^{(i)} = 1 \pm \frac{1}{4} a_i (\cos2\theta+1), 
   \hspace{3mm} a_i=2a_0\pm \frac{1}{2}c_2+a'_{i-1},
\ee
where $\beta=1/c_2\sin^2\theta$ is the turbulent $\beta$ parameter. The expression for $a'_i$ 
is the same as above.

When we assume that the $\theta$-dependent component of $v_\theta''v_\theta''$ equals 
twice that of $v_r''v_r''$, we obtain the same gravitational acceleration as that for 
rotation, Eqs.~(\ref{eq:rgr})-(\ref{eq:rga}), except that $b_D^{(i)}$ is defined in \S\ref{sct:um};
when we assume that the $\theta$-dependent 
component of $v_\theta''v_\theta''$ equals zero, we obtain the same result as that for 
the rotation-like magnetic field, Eqs.~(\ref{eq:hgr})-(\ref{eq:hga}). Therefore,
turbulence plays a role of either rotation or magnetism. The criterion is: we have 
the rotation/magnetism effect when the transverse turbulent velocity is larger/smaller 
than the radial turbulent velocity.

Solar 3D turbulence simulations show that the transverse turbulent velocity is smaller 
than the radial turbulent velocity near the solar surface. We thus expect some 
magnetic effects therein.

\subsubsection{Uniform rotation-magnetism-turbulence equipotential surface}

In the general case, we can express the equipotential surface in the same formula 
as the magnetic equipotential surface:
\be
  x^{(i)} = 1+\frac{1}{4}a_i(\cos2\theta+1),  \label{eq:rmti}
\ee
where
\begin{mathletters}
\ba
a_i &=& a_H\pm 2a_T -a_R + \frac{1}{2}(c_{H2}+ c_{T2})+a'_{i-1},\\
a_R &=& \frac{\Omega_R^2r_e^3}{Gm}, \\
a_T &=& \frac{\Omega_T^2r_e^3}{Gm}, \\
a_H &=& \frac{\Omega_H^2r_e^3}{Gm}, \\
  a_i' &=& \frac{1}{2}[a_i+\frac{r^2}{Gm}(b_D^{(i)}-b_P^{(i)})],\\
  b_P^{(i)} &=& 3\int_0^r \frac{Gma_i}{r^3}dr, \\
  b_D^{(i)} &=& 4\pi G\int_0^r (a_i+c_{H2}+c_{T2})\rho_0 dr.
\ea
\end{mathletters}

Using the equipotential surface profile, Eq.~(\ref{eq:hi}), we can 
calculate the following quantities:
\begin{mathletters}
\ba
  Q^{(i)} &=& 4\pi r_e^2\rho_0\left[1+\frac{1}{3}a_i-\frac{2}{3}(c_{H2}+c_{T2})\right], \\
  \varpi^{(i)} &=& 1+\frac{1}{2}(a_i+c_{H2}+c_{T2})(\cos2\theta+\frac{1}{3}), \\
  \Lambda^{(i)} &=&  -\frac{1}{2}a_i\sin2\theta, \\
  \h{G}_P^{(i)} &=&  \frac{1}{2}\frac{Gm a_i}{r^2}\sin2\theta, \\
  U_P^{(i)} &=& -\frac{1}{2}(\cos2\theta+\frac{1}{3})b_P^{(i)}, \\
  U_D^{(i)} &=& \frac{1}{2}(\cos2\theta+\frac{1}{3})b_D^{(i)}.
\ea
\end{mathletters}
The gravitational acceleration perturbations due to rotation, rotation-like
magnetic field and turbulence are
\begin{mathletters}
\ba
  \h{G}^{(i)} &=& -\frac{1}{2}\left[(\Omega_H^2-\Omega_R^2\pm\Omega_T^2)r
 -\frac{Gm a_i}{r^2}\right]\sin2\theta, \\
  U^{(i)} &=& \frac{1}{2}[3(\Omega_H^2-\Omega_R^2\pm\Omega_T^2)r
+b_D^{(i)}-b_P^{(i)}](\cos2\theta+\frac{1}{3}).
\ea
\end{mathletters}
The gravitational acceleration vector in the system is:
\begin{mathletters}
\ba
  g_r^{(i)} &=& \frac{Gm}{r^2} + \frac{1}{2}[3(\Omega_H^2-\Omega_R^2\pm\Omega_T^2)r
+b_D^{(i)}-b_P^{(i)}](\cos2\theta+\frac{1}{3}), \label{eq:rmtr} \\
  g_\theta^{(i)} &=& -\frac{1}{2}\left[(\Omega_H^2-\Omega_R^2\pm\Omega_T^2)r 
- \frac{1}{2}\frac{Gm a_i}{r^2}\right]\sin2\theta. \label{eq:rmta}
\ea
\end{mathletters}

So far we have assumed that $\Omega_i$ (i = R, H, T) are uniform. They depend
upon $r$ and $\theta$ in general. This is so-called differential rotation. 
We deal with the more complicated situation in the next section.

\subsection{Differential Rotation Rate}

\subsubsection{Differential rotation equipotential surface}

Not all form of differential rotation is non-singular. Whether some
differential rotation is singular is determined by $S_P$, which contains the
term $\h{G}_P\cot\theta$. This term is non-singular if $\h{G}_P$ has a sine
function factor, $\sin\theta$. This criterion yields the following
non-singular differential rotation profile:
\be
  \Omega^2(r,\theta) = \sum_{n=0}^{N}\Omega_{2n}(r)\cos2n\theta, \label{eq:dr}
\ee
where N is an finite integer. This form of expression for $\Omega^2$ is physical because
physical solutions should not be singular.

The first order of approximation to the equipotential surface is
\be
  x^{(0)} = 1  
    -\frac{1}{4}\sum_{n=1}^{N+1} a_{2n}^{(0)}[\cos2n\theta+(-1)^{n-1}],
\ee
where
\Ba
  a_0^{(0)} &=& \frac{1}{4} [2(\overline{\Omega}_0+\Omega_0)+\overline{\Omega}_2-\Omega_2], \\
  a_2^{(0)} &=& \frac{1}{4}\frac{r^3}{Gm}[2(3\overline{\Omega_0}-\Omega_0)
    +2(\overline{\Omega_2}+\Omega_2)-(\overline{\Omega}_4+\Omega_4)], \\
  a_{2n}^{(0)} &=& \frac{1}{4}\frac{r^3}{Gm}\{[(2n+1)\overline{\Omega}_{2n-2}
    -\Omega_{2n-2}]+2(\overline{\Omega}_{2n}+\Omega_{2n})-[(2n-1)\overline{\Omega}_{2n+2}+\Omega_{2n+2}]\}, \\
  U_R &=& -\frac{r}{4}\{(2\overline{\Omega}_0+\overline{\Omega}_2)
+(6\overline{\Omega}_0+2\overline{\Omega}_2-\overline{\Omega}_4)\cos2\theta \nonumber\\
&& +\sum_{n=2}^{N+1}\left[(2n+1)\overline{\Omega}_{2n-2}+2\overline{\Omega}_{2n}
  -(2n-1)\overline{\Omega}_{2n+2}\right]\cos2n\theta\}.
\Ea
We have defined $\overline{\Omega}_0\equiv \frac{1}{r}
\int_0^r\Omega_0dr$, etc.

The next step is to calculate $\varpi^{(0)}$, $U_P^{(0)}$ and $U_D^{(0)}$,
which are used in Eq.~(\ref{eq:x1}). They are
\Ba
\varpi^{(0)} &=& 1 -\frac{1}{2}\sum_{n=1}^{N+1} a_{2n}^{(0)}\left[\cos2n\theta
  +\frac{1}{(2n-1)(2n+1)}\right],\\
\Lambda^{(0)} &=& \frac{1}{2}\sum_{n=1}^{N+1} [n a_{2n}^{(0)}\sin2n\theta],\\
\h{G}_P^{(0)} &=& -\frac{1}{2}\frac{Gm}{r^2}\sum_{n=1}^{N+1}[n a_{2n}^{(0)}\sin2n\theta],\\
U_P^{(0)} &=& \frac{1}{2} \sum_{n=0}^{N+1} b_{P2n}^{(0)}\cos2n\theta,\\
U_D^{(0)} &=& -\frac{1}{2}\sum_{n=0}^{N+1} b_{D2n}^{(0)}\cos2n\theta,
\Ea
where
\Ba
  b_{P0}^{(0)} &=& \int_0^r \frac{Gm}{r^3}\sum_{n=1}^{N+1} n a_{2n}^{(0)}dr, \\
  b_{P2n}^{(0)} &=& \int_0^r\frac{Gm}{r^3}\left[n(2n+1)a_{2n}^{(0)}+\sum_{k=n+1}^{N+1} 2k a_{2k}^{(0)}\right]dr, \\
  b_{D0}^{(0)} &=& 4\pi G\int_0^r \rho\sum_{n=1}^{N+1}\frac{a_{2n}^{(0)}}{(2n-1)(2n+1)}dr,\\
  b_{D2n}^{(0)} &=& 4\pi G\int_0^r \rho a_{2n}^{(0)}dr.
\Ea
The corrected equipotential surface function is
\be
  x^{(i)} = 1  
    -\frac{1}{4}\sum_{n=1}^{N+1} a_{2n}^{(i)}[\cos2n\theta+(-1)^{n-1}],
 \label{eq:drxi}
\ee
where
\begin{mathletters}
\ba
  a_\ell^{(i)} &=& a_\ell^{(0)} + a_\ell^{(i-1)'}, \\
  a_\ell^{(i)'} &=& \frac{1}{2}[a_\ell^{(i)}
    +\frac{r^2}{Gm}(b_{D\ell}^{(i)}-b_{P\ell}^{(i)})],\\
  Q^{(i)} &=& 4\pi r_e^2\rho_0\left[1+\frac{1}{2}\sum_{n=1}^{N+1}\frac{a_{2n}^{(i)}}{(2n-1)(2n+1)}
+\frac{1}{2}\sum_{n=1}^{N+1}(-1)^na_{2n}^{(i)}\right], 
\ea
\end{mathletters}
for $\ell$ = 2, 4, 6, $\cdots$, 2(N+1), and i = 1, 2, 3, $\cdots$

Those terms with $\ell\ne2$ in Eq.~(\ref{eq:drxi}) are 
pure differential rotation effects. The term with $\ell=2$ also contains
some differential rotation correction.

\subsubsection{Differential rotation-like magnetic equipotential surface}

The following toroidal magnetic field mimics the differential rotation, 
Eq.~(\ref{eq:dr}):
\be
B_\phi(r,\theta)=(4\pi\rho)^{1/2}\Omega(r,\theta)r\sin\theta.
\ee
The system has the following equipotential surface:
\be
  x^{(i)} = 1 + \frac{1}{4}\sum_{n=1}^{N+1}a_{2n}^{(i)}\left[\cos2n\theta+(-1)^{n+1}\right],
\ee
where
\begin{mathletters}
\ba
  a_\ell^{(i)} &=& a_\ell^{(0)} + \frac{1}{2}c_\ell + a_\ell^{(i-1)'}, \\
  a_\ell^{(i)'} &=& \frac{1}{2}[a_\ell^{(i)}
    +\frac{r^2}{Gm}(b_{D\ell}^{(i)}-b_{P\ell}^{(i)})],
\ea
\end{mathletters}
for $\ell$ = 2, 4, 6, $\cdots$, 2(N+1), and i = 1, 2, 3, $\cdots$. The starting point $a_\ell^{(0)}$ 
is the same as above except that $U_H=-U_R$.
The coefficients $c_\ell$ are defined by the relation
$\rho=\rho_0/(1-\frac{1}{2}\sum_{n=0}^{N+1} c_{2n}\sin2n\theta)$. They are
\begin{mathletters}
\ba
  c_0 &=& -\frac{\rho_0r^2}{2P_T}\frac{1}{2}(2\Omega_0-\Omega_2), \\
  c_2 &=& \frac{\rho_0r^2}{2P_T}\frac{1}{2}(2\Omega_0-2\Omega_2+\Omega_4), \\
  c_\ell &=& \frac{\rho_0r^2}{2P_T}\frac{1}{2}(\Omega_{\ell-2} -2\Omega_\ell +\Omega_{\ell+2}),
\ea
for $\ell$ = 4, 6, 8, $\cdots$, 2(N+1).
\end{mathletters}

Using these expressions, we can calculate the following quantities:
\Ba
  Q^{(i)} &=& 4\pi r_e^2\rho_0\left[1+\frac{1}{2}c_0-\frac{1}{2}\sum_{n=1}^{N+1}\frac{a_{2n}^{(i)}+c_{2n}}{(2n-1)(2n+1)}-\frac{1}{2}\sum_{n=1}^{N+1}(-1)^{n}a_{2n}\right], \\
\varpi^{(i)} &=& 1 +\frac{1}{2}\sum_{n=1}^{N+1} (a_{2n}^{(i)}+c_{2n})\left[\cos2n\theta
  +\frac{1}{(2n-1)(2n+1)}\right],\\
\Lambda^{(i)} &=& -\frac{1}{2}\sum_{n=1}^{N+1} [n a_{2n}^{(i)}\sin2n\theta],\\
\h{G}_P^{(i)} &=& \frac{1}{2}\frac{Gm}{r^2}\sum_{n=1}^{N+1}[n a_{2n}^{(i)}\sin2n\theta],\\
U_P^{(i)} &=& -\frac{1}{2} \sum_{n=0}^{N+1} b_{P2n}^{(i)}\cos2n\theta,\\
U_D^{(i)} &=& \frac{1}{2}\sum_{n=0}^{N+1} b_{D2n}^{(i)}\cos2n\theta.
\Ea
The coefficients $b_p^{(i)}$ are the same as above, but coefficients $b_D^{(i)}$ are different from
above. They are:
\Ba
  b_{D0}^{(i)} &=& 4\pi G\int_0^r \sum_{n=1}^{N+1}\frac{\rho_0(a_{2n}^{(i)}+c_{2n})dr}{(2n+1)(2n-1)}, \\
  b_{D\ell}^{(i)} &=& 4\pi G\int_0^r \rho_0 (a_\ell^{(i)}+c_\ell)dr 
      \mbox{ for $\ell$ = 2, 4, 6, $\cdots$, 2(N+1)}.
\Ea

\subsubsection{Differential rotation-like turbulent equipotential surface}

The differential rotation-like turbulent parameter is the same as Eq.~(\ref{eq:dr}).
This system has the following equipotential surface in the first approximation:
\be
  x^{(0)} = 1 \mp \frac{1}{4}\sum_{n=1}^{N+1}a_{2n}^{(0)}[\cos2n\theta+(-1)^{n+1}],
\ee
where
\Ba
  a_0^{(0)} &=& \frac{1}{4} [2(\overline{\Omega}_0-\Omega_0)+\overline{\Omega}_2+\Omega_2], \\
  a_2^{(0)} &=& \frac{1}{4}\frac{r^3}{Gm}[2(3\overline{\Omega_0}+\Omega_0)
    +2(\overline{\Omega_2}-\Omega_2)-(\overline{\Omega}_4-\Omega_4)], \\
  a_{2n}^{(0)} &=& \frac{1}{4}\frac{r^3}{Gm}\{[(2n+1)\overline{\Omega}_{2n-2}
    +\Omega_{2n-2})+2(\overline{\Omega}_{2n}-\Omega_{2n})-[(2n-1)\overline{\Omega}_{2n+2}-\Omega_{2n+2}]\}, \\
  U_T &=& \mp\frac{r}{4}\{(2\overline{\Omega}_0+\overline{\Omega}_2)
+(6\overline{\Omega}_0+2\overline{\Omega}_2-\overline{\Omega}_4)\cos2\theta \nonumber\\
&& +\sum_{n=2}^{N+1}\left[(2n+1)\overline{\Omega}_{2n-2}+2\overline{\Omega}_{2n}
  -(2n-1)\overline{\Omega}_{2n+2}\right]\cos2n\theta\}.
\Ea

We have the following recurrence relation:
\be
  x^{(i)} = 1 \mp \frac{1}{4}\sum_{n=1}^{N+1}a_{2n}^{(i)}[\cos2n\theta+(1-)^{n+1}],
\ee
where
\begin{mathletters}
\ba
  a_\ell^{(i)} &=& a_\ell^{(0)} \mp \frac{1}{2} c_\ell + a_\ell^{(i-1)'}, \\
  a_\ell^{(i)'} &=& \frac{1}{2}[a_\ell^{(i)}
    +\frac{r^2}{Gm}(b_{D\ell}^{(i)}-b_{P\ell}^{(i)})],
\ea
\end{mathletters}
for $\ell$ = 2, 4, 6, $\cdots$, 2(N+1), and i = 1, 2, 3, $\cdots$. We can use it to 
express the following quantities:
\Ba
  Q^{(i)} &=& 4\pi r_e^2\rho_0\left[1+\frac{1}{2}c_0\pm\frac{1}{2}\sum_{n=1}^{N+1}\frac{a_{2n}^{(i)}+c_{2n}}{(2n-1)(2n+1)}\pm\frac{1}{2}\sum_{n=1}^{N+1}(-1)^{n}a_{2n}\right], \\
\varpi^{(i)} &=& 1 \mp\frac{1}{2}\sum_{n=1}^{N+1} (a_{2n}^{(i)}+c_{2n})\left[\cos2n\theta
  +\frac{1}{(2n-1)(2n+1)}\right],\\
\Lambda^{(i)} &=& \pm\frac{1}{2}\sum_{n=1}^{N+1} [n a_{2n}^{(i)}\sin2n\theta],\\
\h{G}_P^{(i)} &=& \mp\frac{1}{2}\frac{Gm}{r^2}\sum_{n=1}^{N+1}[n a_{2n}^{(i)}\sin2n\theta],\\
U_P^{(i)} &=& \pm\frac{1}{2} \sum_{n=0}^{N+1} b_{P2n}^{(i)}\cos2n\theta,\\
U_D^{(i)} &=& \mp\frac{1}{2}\sum_{n=0}^{N+1} b_{D2n}^{(i)}\cos2n\theta.
\Ea

\subsubsection{Differential rotation-magnetism-turbulence 
equipotential surface}\label{sct:drmt}

Put all three sources together, we have the following recurrence relation:
\be
  x^{(i)} = 1 + \frac{1}{4}\sum_{n=1}^{N+1}a_{2n}^{(i)}[\cos2n\theta+(-1)^{n+1}],
\ee
where
\begin{mathletters}
\ba
  a_\ell^{(i)} &=& a_{H\ell}^{(0)} \mp a_{T\ell}^{(0)}-a_{R\ell}^{(0)}
       +\frac{1}{2}(c_{H\ell}+ c_{T\ell}) + a_\ell^{(i-1)'}, \\
  a_\ell^{(i)'} &=& \frac{1}{2}[a_\ell^{(i)}
    +\frac{r^2}{Gm}(b_{D\ell}^{(i)}-b_{P\ell}^{(i)})],
\ea
\end{mathletters}
for $\ell$ = 2, 4, 6, $\cdots$, 2(N+1), and i = 1, 2, 3, $\cdots$.
The coefficients $b_p^{(i)}$ are the same as above, but coefficients $b_D^{(i)}$ are:
\Ba
  b_{D0}^{(i)} &=& 4\pi G\int_0^r \sum_{n=1}^{N+1}\frac{\rho_0(a_{2n}^{(i)}+c_{H2n}+c_{T2n})dr}{(2n+1)(2n-1)}, \\
  b_{D\ell}^{(i)} &=& 4\pi G\int_0^r \rho_0 (a_\ell^{(i)}+c_{H\ell}+c_{T\ell})dr 
      \mbox{ for $\ell$ = 2, 4, 6, $\cdots$, 2(N+1)}.
\Ea

The useful quantities are:
\Ba
  Q^{(i)} &=& 4\pi r_e^2\rho_0\left[1+\frac{1}{2}(c_{H0}+c_{T0})-\frac{1}{2}\sum_{n=1}^{N+1}\frac{a_{2n}^{(i)}+c_{H2n}+c_{T2n}}{(2n-1)(2n+1)}-\frac{1}{2}\sum_{n=1}^{N+1}(-1)^{n}a_{2n}\right], \\
\varpi^{(i)} &=& 1 +\frac{1}{2}\sum_{n=1}^{N+1} (a_{2n}^{(i)}+c_{H2n}+c_{T2n})\left[\cos2n\theta
  +\frac{1}{(2n-1)(2n+1)}\right],\\
\Lambda^{(i)} &=& -\frac{1}{2}\sum_{n=1}^{N+1} [n a_{2n}^{(i)}\sin2n\theta],\\
\h{G}_P^{(i)} &=& \frac{1}{2}\frac{Gm}{r^2}\sum_{n=1}^{N+1}[n a_{2n}^{(i)}\sin2n\theta],\\
U_P^{(i)} &=& -\frac{1}{2} \sum_{n=0}^{N+1} b_{P2n}^{(i)}\cos2n\theta,\\
U_D^{(i)} &=& \frac{1}{2}\sum_{n=0}^{N+1} b_{D2n}^{(i)}\cos2n\theta, \\
  g_r^{(i)} &=& \frac{Gm}{r^2} + U_H + U_T + U_R
+ \frac{1}{2}\sum_{n=0}^{N+1}(b_{D2n}^{(i)}-b_{P2n}^{(i)})\cos2n\theta, \\
  g_\theta^{(i)} &=& -\frac{1}{2}(\Omega_H^2\mp\Omega_T^2-\Omega_R^2)r\sin2\theta
  +\frac{1}{2} \frac{Gm}{r^2}\sum_{n=1}^{N+1}na_{2n}^{(i)}\sin2n\theta.
\Ea

This is the general case for rotation, the rotation-like toroidal magnetic field 
and turbulence. The recurrence relations given here reflect the real cause-effect
relation. The source terms $(U_R-\h{R}_r)$, $(U_H-\h{H}_r)$ and $(U_T-\h{T}_r)$ are
the causes, and $U_P$, $U_D$ and $\varpi$ are their effects. When some asphericity
sources are present, the spherically-symmetric star should readjust to assume an
aspherical equilibrium configuration. The recurrence relations describe the
readjustment procedure.

\section{METHOD OF SOLUTION}\label{sct:s4}

\subsection{2D Stellar Structure Equations with an Known Equipotential Surface}\label{sct:stellar}

For the cases studied above, we can use the recurrence relations to 
calculate the equipotential surface functions $x^{(i)}$ to certain accuracy. 
The result is denoted as $x=x^{(\infty)}$. From now on, we use the un-superscripted 
symbols to express the corresponding limits, for example, $a_\ell=a_\ell^{(\infty)}$, 
and so on. We then use $x$ to calculate functions $\varpi$, $\Lambda$, $Q$, etc. 
This is equivalent to solving the Poisson equation for the gravitational acceleration vector.

With the help of the equipotential surface, what we need to numerically solve for 
are $r_e$, $P_T$, $T$, and $L$, which are governed by the following four equations:
\begin{mathletters}
\ba
  \p{r_e'}{s} &=& \frac{m}{Q r_e}, \label{eq:st1a} \\
  \p{P'}{s} &=& -\frac{Gm^2}{4\pi r_e^4 P_T}, \label{eq:st2a} \\
\p{T'}{s} &=& \p{P'}{s} \left\{ \begin{array}{ll}
         \nabla_{\s{rad}} & \mbox{ radiative} \\
         \nabla_{\s{c}} & \mbox{ convective} \\
    \end{array} \right. \label{eq:st3a} \\
\p{L}{s} &=& \frac{m\varpi}{L_\sun}\left(\epsilon-T\od{S_T}{t}\right)
  -\frac{m \varpi\Psi}{L_\sun r_e\rho}. \label{eq:st4a}
\ea
\end{mathletters}
Here $r'_e=\ln r_e$, $r=r_ex$, $\varpi=\rho/\rho_m$, 
$\Lambda=(\partial x/\partial\theta)_m$, and
\begin{mathletters}
\ba
\Psi &=& F_\theta\cot\theta+\p{F_\theta}{\theta}, \\
F_\theta &=&  \tilde{P}(F^1+F^2+F^3), \\
\tilde{P} &=& \frac{G m Qx\Lambda}{4\pi r_e^3 P_T},\\
F^1 &=& -\frac{4ac T^4}{3\kappa\rho}\frac{\nabla}{r}, \\
F^2 &=& -\frac{1}{2} \frac{\rho C_PT l_m v_{\s{conv}}}{1+v_{\s{conv}}/v_0}\frac{\nabla}{r}, \\
F^3 &=& \frac{1}{2} \frac{\rho C_P T l_m v_{\s{conv}}}{1+v_{\s{conv}}/v_0}\frac{\nabla'_{\s{ad}}}{r}.
\ea
\end{mathletters}

The variable $\Psi$ has a term that is proportional to the following expression:
\ba
 \tilde{\Lambda} & \equiv & \Lambda\cot\theta+\p{\Lambda}{\theta} \nonumber \\
  &=& -\frac{1}{2}\sum_{n=1}^{N+1}na_{2n}-\frac{1}{2}\sum_{n=1}^{N+1}\left[n(2n+1)a_{2n}
+\sum_{k=n+1}^{N+1}2k a_{2k}\right]\cos2n\theta. \label{eq:tlambda}
\ea
The second term is $\h{F}\tilde{P}^2$, where $\h{F}$ is defined in \S\ref{sct:upd}. 
The required $\Psi$ is the sum of these two terms:
\be
  \Psi = \frac{G m Q\tilde{\Lambda}}{4\pi r_e^3 P_T}(F^1+F^2+F^3)+\h{F}\tilde{P}^2.
\ee
The other supplement quantities are given in \S\ref{sct:drmt}. 

\subsection{Linearization of 2D Stellar Structure Equations}

The construction of a two-dimensional stellar model begins 
by dividing the star into $M$ mass shells and $N$ angular 
zones. The mass shells are assigned a value $s_i=\log m_i$, 
where $m_i$ is the interior mass at the midpoint of shell i. 
The angular zones are assigned a value $\theta_j$. A starting 
(or previous in evolutionary time) model is supplied with a 
run of ($P'_{i}$, $T'_{ij}$, $r'_{i}$, $L_{ij}$,
$U_{ij}=0$, $\h{G}_{ij}=0$) for i=1 to $M$ and j=1 to $N$. 

Different terms in Eqs.~(\ref{eq:st1a})-(\ref{eq:st4a}) have different 
derivatives with respect to the stellar parameters ($P_T$, $T$, $r$, 
$L$). These derivatives are needed to write down 
the linearized difference equations. We hence rewrite them as follows:
\begin{mathletters}
\ba
\p{P'}{s} &=& \h{P}, \label{eq:pdiff1} \\
\p{T'}{s} &=& \h{T}, \label{eq:pdiff2} \\
\p{r'}{s} &=& \h{R}, \label{eq:pdiff3} \\
\p{L}{s} &=& \sum_{\ell=1}^3 \h{L}^\ell. \label{eq:pdiff4}
\ea
\end{mathletters}
The symbols used above are defined as follows:
\begin{mathletters}
\ba
\h{P} &\equiv& -\frac{Gm^2}{4\pi r_e^4 P_T}, \\
\h{T} &\equiv& \h{P}\nabla, \\
\h{R}   &\equiv& \frac{m}{Q r_e}, \\
\h{L}^1 &\equiv& \frac{m\varpi}{L_\sun}\left(\epsilon-T\od{S_T}{t}\right), \\
\h{L}^2 &\equiv& -\frac{m\varpi}{L_\sun r\rho}F_\theta\cot\theta, \\
\h{L}^3 &\equiv& -\frac{m\varpi}{L_\sun r\rho}\p{F_\theta}{\theta}.
\ea
\end{mathletters}

We use the central difference scheme to 
approximate the stellar structure equations. 
The corresponding difference equations are
\begin{mathletters}
\ba
F^i_P &\equiv& (P'_i-P'_{i-1}) - \frac{1}{2}\Delta s_i
  (\h{P}_{i}+\h{P}_{i-1})=0, \label{eq:diff1}\\
F^{ij}_T &\equiv& (T'_{ij}-T'_{i-1j}) - \frac{1}{2}\Delta s_i
  (\h{T}_{ij}+\h{T}_{i-1j})=0, \label{eq:diff2}\\
F^{i}_R &\equiv& (R'_{i}-R'_{i-1}) - \frac{1}{2}\Delta s_i
  (\h{R}_{i}+\h{R}_{i-1})=0, \label{eq:diff3} \\
F^{ij}_L &\equiv& (L_{ij}-L_{i-1j}) - \frac{1}{2}\Delta s_i
  \sum_{\ell=1}^3 (\h{L}^\ell_{ij}+\h{L}^\ell_{i-1j})=0, \label{eq:diff4}
\ea 
\end{mathletters}
for i = 2 to M, and j = 1 to N.
The linearization of Eqs.~(\ref{eq:diff1})-(\ref{eq:diff4}) with respect 
to ($\delta P'_{ij}$, $\delta T'_{ij}$, $\delta r'_{ij}$, and 
$\delta L_{ij}$) yields $2(M-1)N+2(M-1)$
equations for the $2MN+2M$ unknowns. The $N+1$ additional 
equations are supplied by the boundary conditions at the center:
\begin{mathletters}
\ba
  F^{1}_R &\equiv & r'_1-[s_1-\ln(Q/3)] =0, \label{eq:center1} \\
  F^{1j}_L &\equiv & L_{1j}- \sum_{\ell=1}^3 \h{L}^\ell_{1j}=0, \label{eq:center2}
\ea
\end{mathletters}
where j = 1 to $N$. Another $N+1$ additional equations are 
supplied by the boundary conditions at the surface:
\begin{mathletters}
\ba
  F^{M+1}_R &\equiv & R'_{M} - a_1 P'_{M} - a_2 T'_{MN} - a_3=0, \label{eq:surf1} \\
  F^{M+1j}_L &\equiv & L_{Mj}' (\ln L_{Mj}- a_4 P'_{M} 
    - a_5 T'_{Mj} - a_6)=0, \label{eq:surf2}
\ea
\end{mathletters}
where j = 1 to $N$. The F equations are linearized,
\begin{mathletters}
\ba
   -F^{ij}_w &=& \sum_{l=1}^{M}\sum_{k=1}^{N}\left(\p{F^{ij}_w}{R'_{l}}
  \delta R'_{l} 
+\p{F^{ij}_w}{L_{lk}}\delta L_{lk}
 +\p{F^{ij}_w}{P'_{l}}\delta P'_{l} 
+\p{F^{ij}_w}{T'_{lk}}\delta T'_{lk}\right)  \nonumber \\
 &&  \mbox{ for w = T, L},  \label{eq:linear} \\
 -F^{i}_w &=&  \sum_{l=1}^{M}\left(\p{F^{i}_w}{R'_{l}}\delta R'_{l} 
 +\p{F^{i}_w}{P'_{l}}\delta P'_{l} 
\right)  
  \mbox{ for w = R, P }, \label{eq:linear2}
\ea
\end{mathletters}
where $i = 1$ to $M$; and j = 1 to $N$. 
The summation over l has non-zero terms only for l = i-1, i;
the summation over k has non-zero terms only for k = j.
See appendix A for the coefficient matrix elements.

Since we explicitly take advantage of the equipotential surface function $x$,
we can express the derivatives of all dependent variables with respect 
to $\theta$ in terms of $\Lambda$, which is the $\theta$-derivative of $x$ 
on the equipotential surface. This unchains the explicit binding between adjacent 
angular zones and allows us to treat each zone as if it is a 
one-dimensional problem. However, the implicit binding cannot be broken 
because of the mass conservation requirement that is characterized by the 
parameter $Q$, which is an integral over all zones.

These equations can be solved by means of the Henyey method.

\subsection{Non-equator Reference Surface}

So far we have used the equator as the reference surface. This is not necessary. We can 
use the other reference surface instead, say, $\theta=\theta_0$. The equator is only a 
specific example where $\theta_0=\pi/2$. We need a non-equator reference surface when 
the applied field peaks at or near the equator. We use the subscript "f" as the indicator 
of the reference surface $\theta=\theta_0$. Since $r=r_f x$, 
the equipotential surface $x=x(r_f,\theta)$ should be normalized to unity at the 
reference surface $\theta=\theta_0$. We give different formulas as follows:
\begin{mathletters}
\ba
  x^{(i)} &=& 1 + \frac{1}{4}\sum_{n=1}^{N+1} a_{2n}^{(i)}[\cos2n\theta-\cos2n\theta_0], \\
 Q^{(i)} &=& 4\pi r_e^2\rho_0\left\{1+\frac{1}{2}(c_{H0}+c_{T0})-\frac{1}{2}\sum_{n=1}^{N+1}
\left[\frac{a_{2n}^{(i)}+c_{H2n}+c_{T2n}}{(2n-1)(2n+1)}+a_{2n}^{(i)}\cos2n\theta_0\right]\right\}.
\ea
\end{mathletters}
We use subscript "f" to replace "e" in the other formulas and/or equations.

\section{HIGH-PRECISION 2D SOLAR MODELS}\label{sct:s5}

The solar variability models need to be accurate enough to match the 
seismic structures of the Sun (Gough et al 1996), as the (1D) 
standard solar models do (Bahcall et al 2006 and 
references cited therein). Standard solar models (1) use the most 
accurate available input parameters, including radiative opacity, 
equation of state, and nuclear cross sections, (2) include element diffusion,
and (3) have a high numerical resolution. Our 2D models inherit all these 
features because our 2D code described in this paper is a natural extension 
of YREC (Yale Rotation Evolution Code) to two dimensions. We also tested its 
1D counterpart with turbulence (Li et al 2002) and made sure 
that the resultant 1D solar models are accurate enough to meet with 
our accuracy requirements. We further tested the 1D code with magnetic 
fields and turbulence (Li et al 2003) to make sure that it is accurate enough 
to discern the solar cycle-related p-mode frequency changes. These demonstrate 
that the first dimension is accurate enough to discern the solar 
cycle-related changes. The number of mass layers used in both 1D and 2D model 
calculations is more than 2500.

\subsection{Error Controls}

Here we describe how we control the numerical errors to meet with our 
accuracy requirement.

\subsubsection{Radial}

This is the same as its 1D counterpart. The numerical errors are controlled in terms 
of two parameters $\epsilon_F$ and $\epsilon_C$:
\be
   |F_w^{ij}| < \epsilon_F, \mbox{ and } |\delta w^{ij}| < \epsilon_C
\ee
for i = 1 to M+1, j = 1 to N, and w = P, T, R, and L. See Eqs.~(\ref{eq:diff1}-\ref{eq:surf2})
for the definition of $F_w^{ij}$.

The 1D standard solar models have $\epsilon_F\sim \epsilon_C \sim 10^{-6}$, which is the 
relative accuracy of the numerical solution of the stellar structure equations.
We use the same values of $\epsilon$ for our 2D solar models.

\subsubsection{Colatitudinal}

From \S\ref{sct:stellar} we can see that the colatitudinal factors affect 
the stellar structure equations in terms of $x$, $Q$, $\varpi$, $\Lambda$, 
and $\tilde{\Lambda}$. The quantities $Q$ and $\varpi$ are the integrals of $x$ 
over $\theta$, and $\Lambda$ and $\tilde{\Lambda}$ are the (first-order and second-order) 
derivatives of $x$ with respect to $\theta$. Therefore, the colatitudinal errors are 
determined by the error of the equipotential surface function $x$, which is defined by 
Eq.~(\ref{eq:x1}).

For rotation, rotation-like magnetic field, and/or rotation-like turbulence that 
are symmetric with respect to the equator, Eq.~(\ref{eq:x1}) can be rewritten as follows:
\be
  x = q^{-1/2}\left(1-\frac{1}{2}\sum_{n=0}^{N+1}c_{2n}\cos2n\theta\right)^{-1/2}\left(1+x^3\sum_{n=0}^{\infty}u_{2n}\cos2n\theta\right)^{1/2}. \label{eq:it1}
\ee 
In doing so we have rewritten $\rho_m$ (Eq.~\ref{eq:rhom1}) and $\rho$ as follows:
\Ba
  \rho_m &=& \rho_0 x^{-2}q, \\
  \rho &=& \rho_0 \left(1-\frac{1}{2}\sum_{n=0}^{N+1}c_{2n}\cos2n\theta\right)^{-1}.
\Ea
We have also used the fact that $U\propto x$, $\h{H}_r\propto x$, 
$\h{T}_r\propto x$, and $\h{R}_r\propto x$. The quantities used here are defined as follows:
\begin{mathletters}
\ba
  q &=& \int_0^{\pi/2}\left(1-\sum_{n=0}^{N+1}c_{2n}\cos2n\theta\right)^{-1}x^2\sin\theta d\theta, 
    \label{eq:it2} \\
  u_{2n} &=& a_{2n}^{(0)} + \frac{1}{2}\frac{r_e^3}{Gm}(b_{D2n}-b_{P2n}), \label{eq:it3} \\
  b_{D0} &=& 4\pi G\int_0^r \sum_{n=1}^{\infty}\frac{\rho_0(a_{2n}+c_{2n})dr}{(2n+1)(2n-1)}, 
    \label{eq:it4} \\
  b_{D2n} &=& 4\pi G\int_0^r \rho_0 (a_{2n}+c_{2n})dr, \label{eq:it5} \\
  b_{P0} &=& \int_0^r \frac{Gm}{r_e^3}\sum_{n=1}^{\infty} n a_{2n}dr, \label{eq:it6} \\
  b_{P2n} &=& \int_0^r\frac{Gm}{r_e^3}\left[n(2n+1)a_{2n}+\sum_{k=n+1}^{\infty} 2k a_{2k}\right]dr,
    \label{eq:it7}
\ea
\end{mathletters}
for n = 1 to $\infty$. Here we have used the Fourier series to express the normalized equipotential
surface function $x$:
\be
  x = 1 + \sum_{n=1}^{\infty} a_{2n}[\cos 2 n \theta+(-1)^{n+1}]. \label{eq:it8}
\ee
For pure rotation, $c_{2n}=0$ for all n (n = 0 to $\infty$).

In practice, we have to truncate the infinite Fourier series to approximate $x$,
\be
  x_{\h{N}} = 1 + \sum_{n=1}^{\h{N}}[\cos2n\theta+(-1)^{n+1}]. \label{eq:it9}
\ee
Since $|\cos2n\theta|\le 1$, the truncation error can be estimated as follows:
\be
  \epsilon_x \equiv |x-x_\h{N}| \le \sum_{n=\h{N}+1}^\infty|a_{2n}|. \label{eq:it10}
\ee
If the $a_{2n}$'s are rapidly decreasing, which is the typical case, then the 
truncation error is dominated by $a_{2(\h{N}+1)}$. We can thus use $a_{2(\h{N}+1)}$ 
as an estimate of the truncation error of $x$:
\be
  \epsilon_x \sim |a_{2(\h{N}+1)}|. \label{eq:it11}
\ee

We want to achieve a relative accuracy of $10^{-6}$ for the stellar 
parameters P, T, R and L in the 2D model, the same as in the 1D standard solar model.
This requires the similar relative accuracy for $x$. Since $x$ 
is of the order of magnitude of unity, its relative error is the same as 
its absolute error. In order to achieve such high an accuracy, we use three-level 
iterations to solve Eqs.~(\ref{eq:it1}-\ref{eq:it8})

The first-level iteration is given in \S\ref{sct:s3} in terms of the recurrence 
relations, which are based on the linear approximation of Eq.~(\ref{eq:x1}).
The convergence criterion is $|a^{(i)}_{2n}-a^{(i-1)}_{2n}| < \epsilon$ 
for i = 1 to N+1, where $\epsilon=10^{-6}$. The converged $a^{(i)}_{2n}$'s are 
denoted by $a^I_{2n}$. The second- and third-level iterations are used to do 
nonlinear corrections.

The second-level iteration uses
\be
  x^{(0)}_{II} = 1 + \sum_{n=1}^{\h{N}+1}a^I_{2n}[\cos2n\theta+(-1)^{n+1}]
\ee
as the initial guess for $x$ in Eq.~(\ref{eq:it1}). The updated $x_{II}^{(i)}$ is normalized 
as follows:
\be
  x_{II}^{(i)} = x_{II}^{(i)}-x_{II}^{(i)}(\theta=\pi/2) + 1
\ee
for i = 1, 2, 3, $\cdots$. The convergence criterion is
\be
  |x_{II}^{(i)}-x_{II}^{(i-1)}| < \epsilon.
\ee
The converged $x_{II}^{(i)}$ is denoted by $x_{II}$, which is then expanded as the Fourier series
to prepare for the third-level iteration:
\be
  x_{II} = \sum_{n=0}^{\infty} a^{II}_{2n} \cos2n\theta. \label{eq:xii}
\ee
We have to truncate Eq.~(\ref{eq:xii}) to go further. The truncation criterion is
\be
  |a_{2\h{N}}|\ge \epsilon \mbox{ and } |a_{2n}^{II}| < \epsilon \mbox{ for $n\ge \h{N}+1$}.
\ee
Generally speaking, $\h{N} \ge N+1$.

Using $a^{II}_{2n}$ (n = 1 to $\h{N}$) as the initial guess for $a_{2n}^{III}$, denoted as 
$b_{2n}^{(0)}$, we repeat the second-level iteration to update $b_{2n}^{(i)}$.
The convergence criterion is
\be
   |b_{2n}^{(i)}-b_{2n}^{(i-1)}| < \epsilon
\ee
for n = 1 to $\h{N}$. The converged $b_{2n}^{(i)}$'s are denoted as $a_{2n}^{III}$.
Using $a_{2n}^{III}$, we can calculate $x$, $Q$, $\varpi$, $\Lambda$, 
$\tilde{\Lambda}$, and other quantities such as $g_r$ and $g_\theta$.

Extensive numerical experiments reveal that the dominant error sources 
come from Eq.~(\ref{eq:it7}), whose integrand is proportional to the Fourier expansion 
coefficients of $\tilde{\Lambda}$, Eq.~(\ref{eq:tlambda}):
\be
  \tilde{\Lambda}_{2n} = n(2n+1)a_{2n}+\sum_{k=n+1}^{\h{N}+1}2ka_{2k}  
    \mbox{ for n = 1 to $\h{N}$+1}.
\ee
Its first term originates from the second derivative of the equipotential surface $x$. 
The coefficient $n(2n+1)$ of $a_{2n}$ in the first term will substantially magnify the 
error of $a_{2n}$ when $n$ is big. In order to control this error, we calculate the maximal 
value of the ratio of the centrifugal over the gravitational acceleration for pure rotation, 
denoted as $\eta$, we define $\eta$ as the maximal value of $1/\beta$ for magnetic fields 
and/or turbulence. Numerical experiments show that the convergence criterion is
$\epsilon=\max(\epsilon_F,\epsilon_C,\eta^5)$.

\subsection{Examples}

\subsubsection{Uniform rotation}

\begin{figure}
\plotone{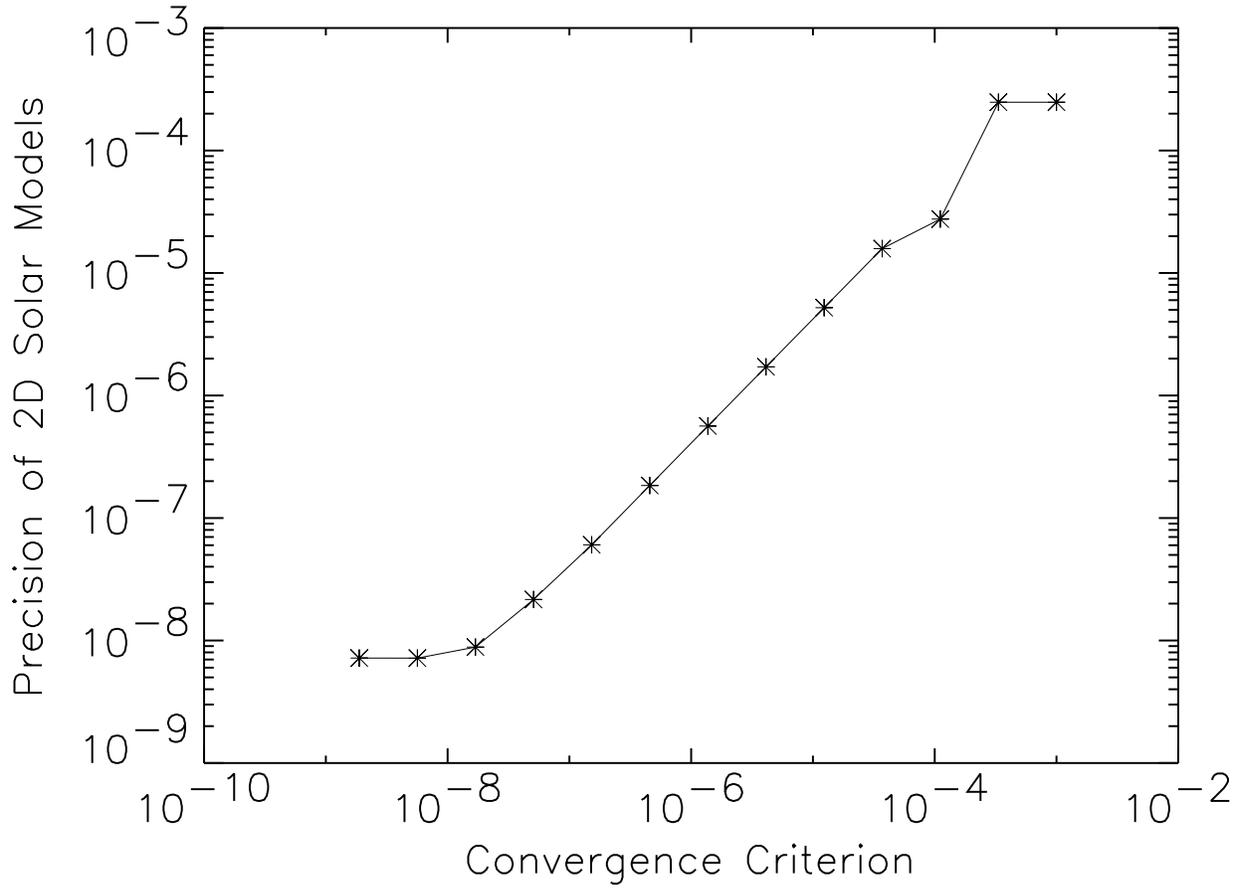}
\caption{The numerical accuracy of the 2D solar models with a zero-rotation rate as a function of
convergence criteria. The symbols mark the data points.}\label{fig:eps0}
\end{figure}

This is the simplest case. First of all we calculate a high-precision (1D) standard solar model 
by using the convergence criterion $\epsilon_F=\epsilon_C=1\times10^{-10}$. We use it as the 
benchmark. We then use zero-rotation rate ($\Omega=0$) to calculate a series of 2D solar models by
using the convergence criterion $\epsilon=\epsilon_F=\epsilon_C$ from $1\times10^{-3}$ 
to $1\times10^{-9}$. The numerical accuracy of the 2D solar models is measured in terms of 
their relative errors with respect to the standard solar model. The model is represented in
terms of runs of pressure, $P=P(m,\theta)$, temperature $T=T(m,\theta)$, radius $r=r(m,\theta)$,
luminosity $L=L(m,\theta)$, and density $\rho=\rho(m,\theta)$. The numerical accuracy of the 2D
solar models is thus defined as the maximal value of the relative errors for all five 
variables over all grid points. The results are shown in Fig.~\ref{fig:eps0}, in which the symbols
mark the data points. The figure shows that we can achieve a precision significantly better than
$1\times10^{-6}$, which is accurate enough for the relevant solar applications. Since we avoid
numerical derivatives and integrals, the results are independent of the grid size in the second
coordinate $\theta$. This is confirmed by the detailed model calculations by setting N = 9, 17, and
33, where N is the number of grid points in the second dimension. For both 1D and 2D models the first
dimension has the same grid point number M = 2576.

\begin{figure}
\plotone{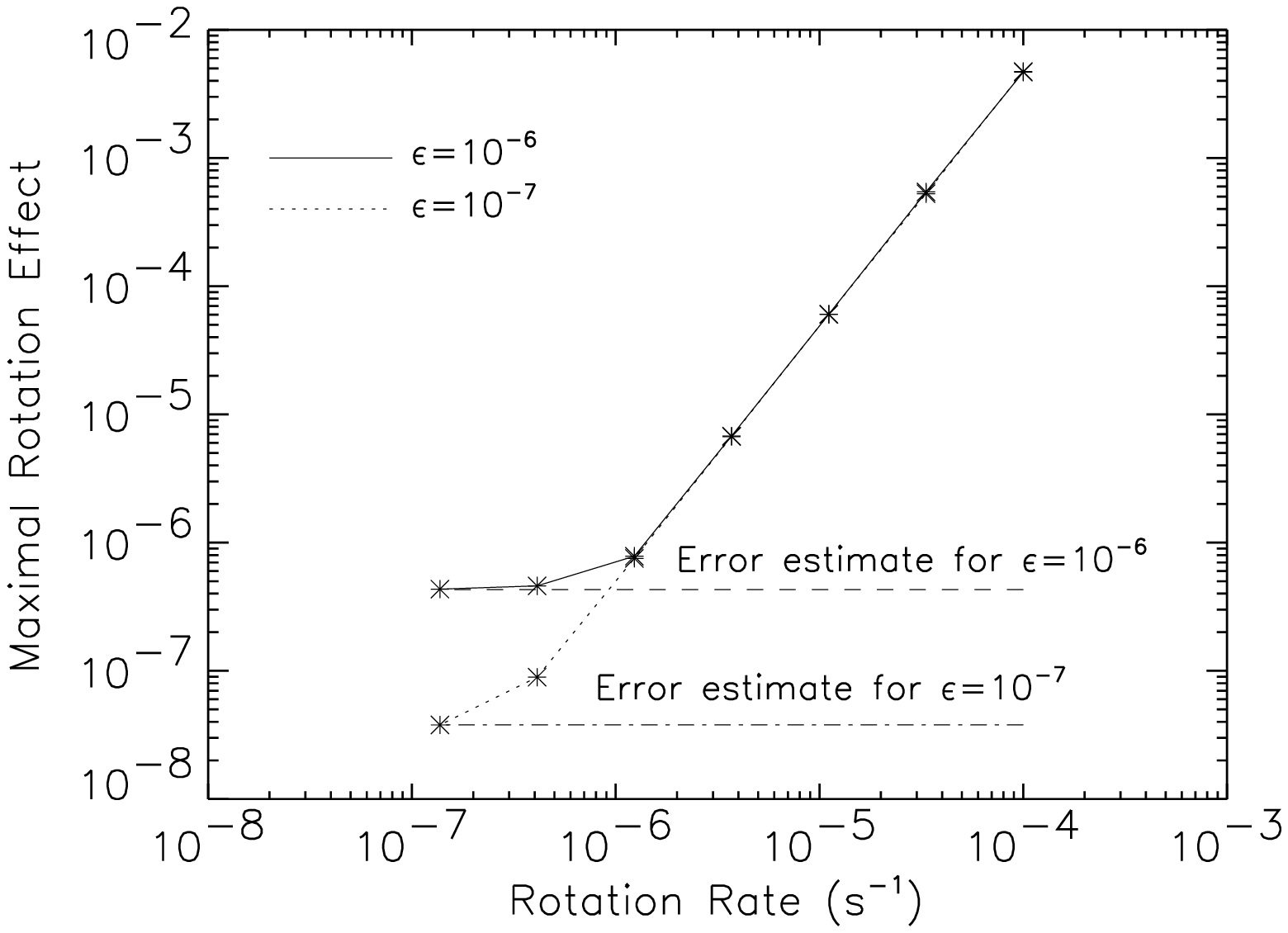}
\caption{Maximal rotation effect as a function of the rotation rate $\Omega$. The symbols mark the data points, and the dashed line shows the relative error estimate.}\label{fig:eps1}
\end{figure}

When the rotation rate is nonzero, i.e., $\Omega\ne0$, the relative differences between the 2D and 
1D models such as $\h{E}_P=[P(m,\theta)-P(m)]/P(m)$ etc can be considered to be the rotation effects.
They are functions of the rotation rate $\Omega$, convergence criterion $\epsilon$, the mass
coordinate $m$ and colatitude coordinate $\theta$, for example,
$\h{E}_P=\h{E}_P(m,\theta;\Omega,\epsilon)$, $\h{E}_T=\h{E}_T(m,\theta;\Omega,\epsilon)$, and
similar expressions for $r$,  $L$ and $\rho$. Their accuracy is estimated by the corresponding
value at the zero-rotation rate. Fig.~\ref{fig:eps1} shows how the maximal value of $\h{E}_P$,
$\h{E}_T$, $\h{E}_r$, $\h{E}_L$, and $\h{E}_\rho$ changes with $\Omega$, where we fix
$\epsilon=1\times10^{-6}$ (solid line) or $1\times10^{-7}$ (dotted line). 
So the relative error is of the same order as $\epsilon$, as indicated by the dashed line 
($\epsilon=1\times10^{-6}$) and the dot-dashed line ($\epsilon=1\times10^{-7}$) in the figure.

\begin{figure}
\plotone{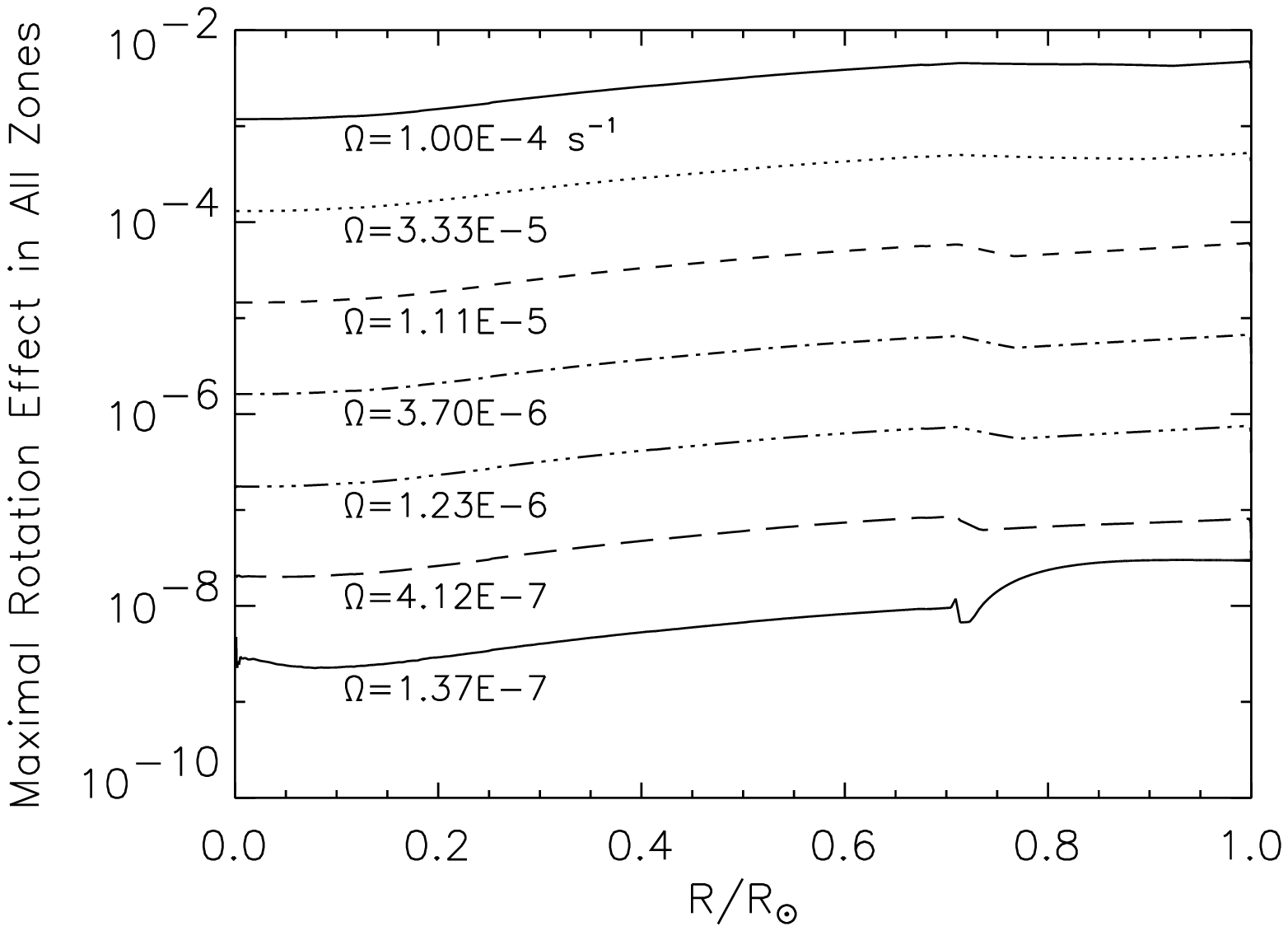}
\caption{Maximal rotation effect in all zones as a function of both the rotation rate $\Omega$
and radius $R/R_\sun$.}\label{fig:eps2}
\end{figure}

\begin{figure}
\plotone{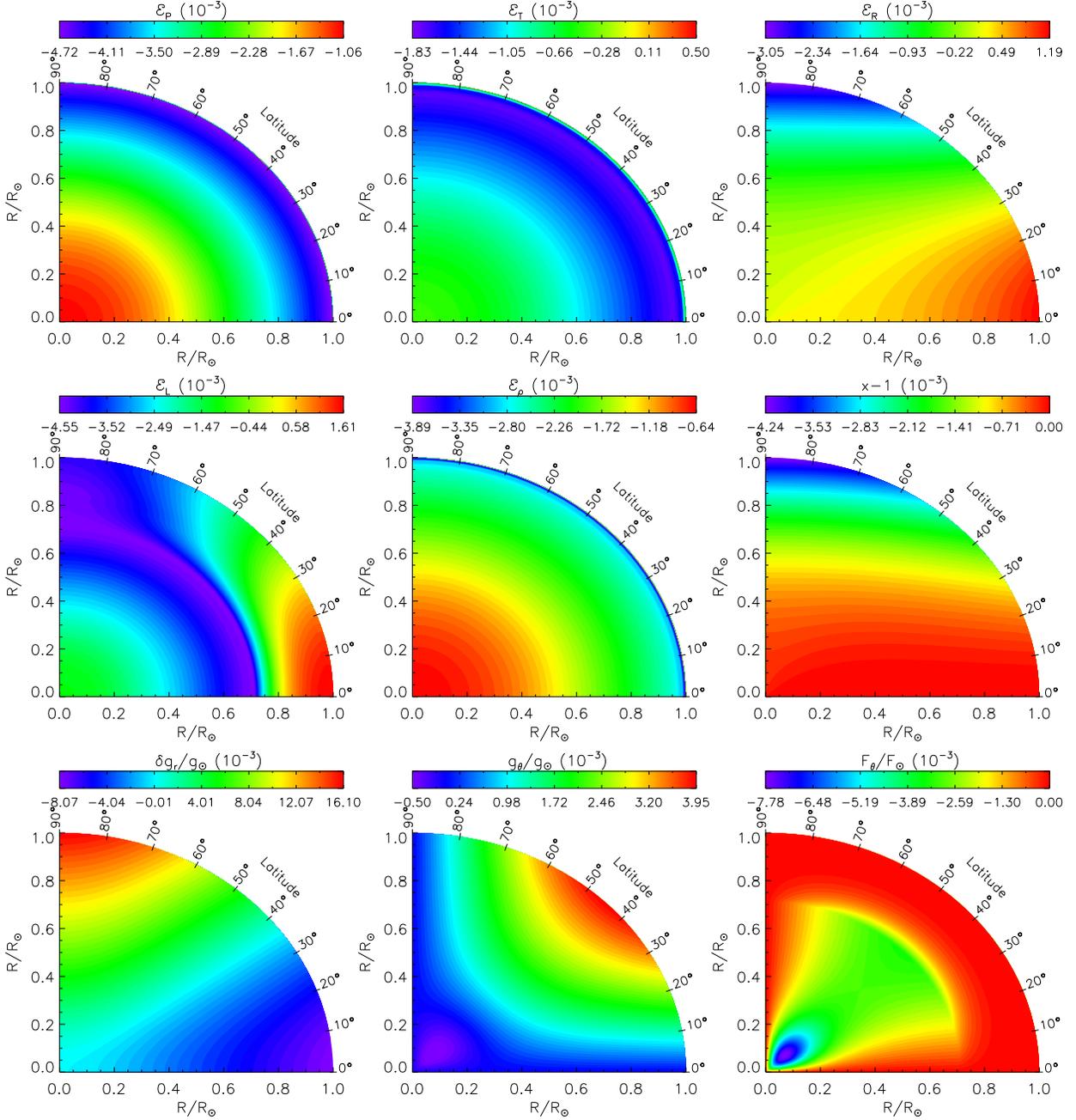}
\caption{Contours of the 2D solar model with a uniform rotation rate
$\Omega=10^{-4}$ s$^{-1}$. The top five sub-figures show the detail dependence of $\h{E}_P$
to $\h{E}_\rho$ on $R/R_\sun$ and $\theta$. The last fore sub-figures show the equipotential 
surface function $x-1$, the transverse component of energy flux $\vb{F}$, $F_\theta/F_\sun$, 
the radial perturbation component, and transverse component of the gravitational acceleration.}
\label{fig:rot}
\end{figure}

To see where the maximal rotation effect takes place, we plot 
$\h{E}_R=\h{E}(m(R), \{\theta\}; \Omega)$ as a function of $R/R_\sun$ and $\Omega$ in 
Fig.~\ref{fig:eps2}, where $\h{E}_R$ 
is the maximal value among $\h{E}_P$, $\h{E}_T$, $\h{E}_r$, $\h{E}_L$ and $\h{E}_\rho$ over all
zones, and $R$ is the radius of the mass shell $m$ in the standard solar model. Similarly, we have
$\h{E}_\theta=\h{E}(\{m\}, \theta; \Omega)$. Since it changes little with $\theta$, we do not need
to plot it. Fig.~\ref{fig:eps2} shows that the maximum takes place at the base 
of the convection zone or near the surface. Fig.~\ref{fig:rot} shows the detail dependence of
$\h{E}_P$, $\h{E}_T$, $\h{E}_r$, $\h{E}_L$ and $\h{E}_\rho$ on $R/R_\sun$ and $\theta$. It 
also shows the equipotential surface $x$, $F_\theta$, $\delta g_r$ and $g_\theta$, which have no 1D
counterparts.

\subsubsection{Uniform rotation-like magnetic field}

\begin{figure}
\plotone{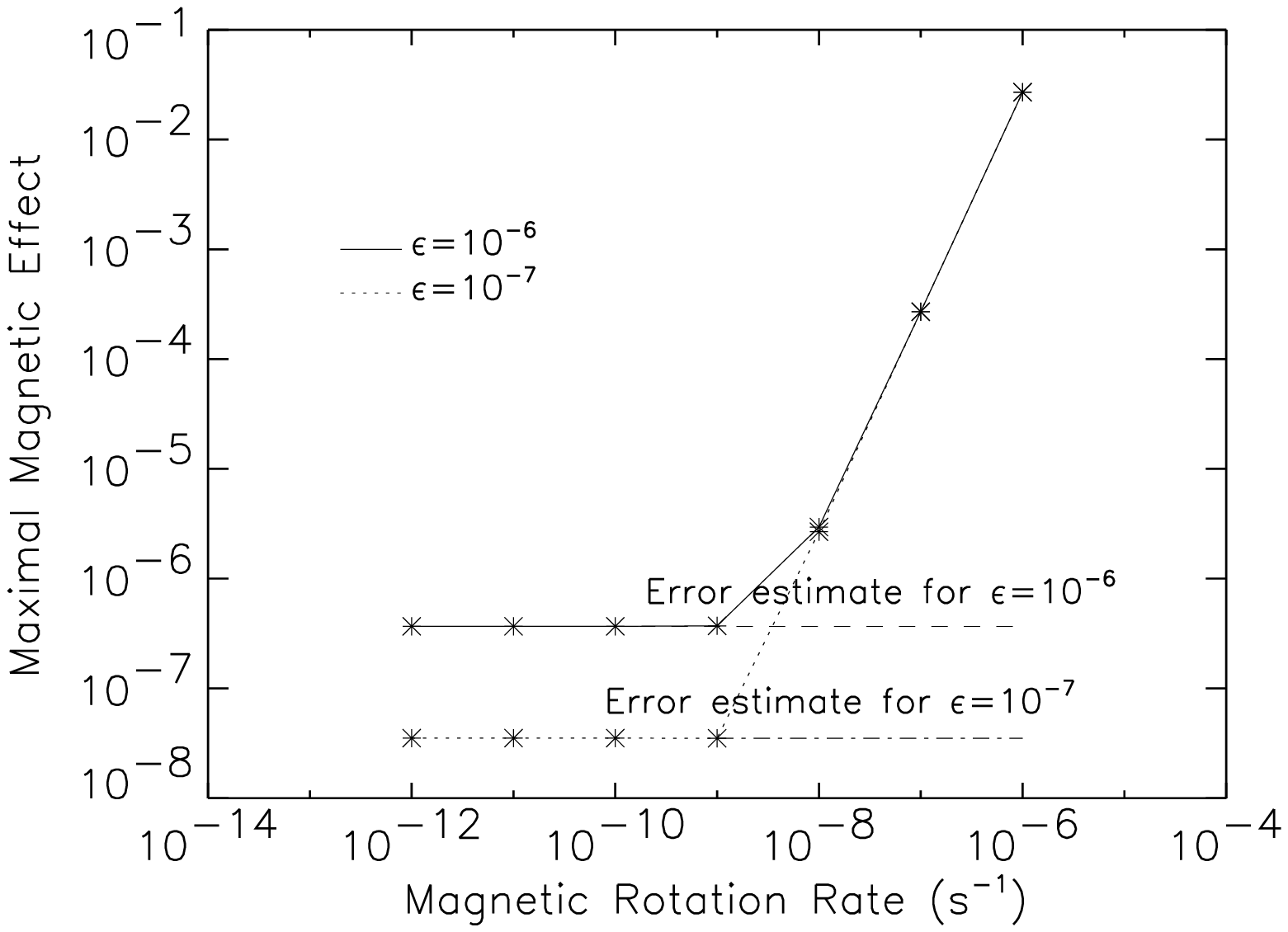}
\caption{Maximal rotation effect as a function of the rotation rate $\Omega$. The symbols mark the data points, and the dashed line shows the relative error estimate.}\label{fig:eps3}
\end{figure}

\begin{figure}
\plotone{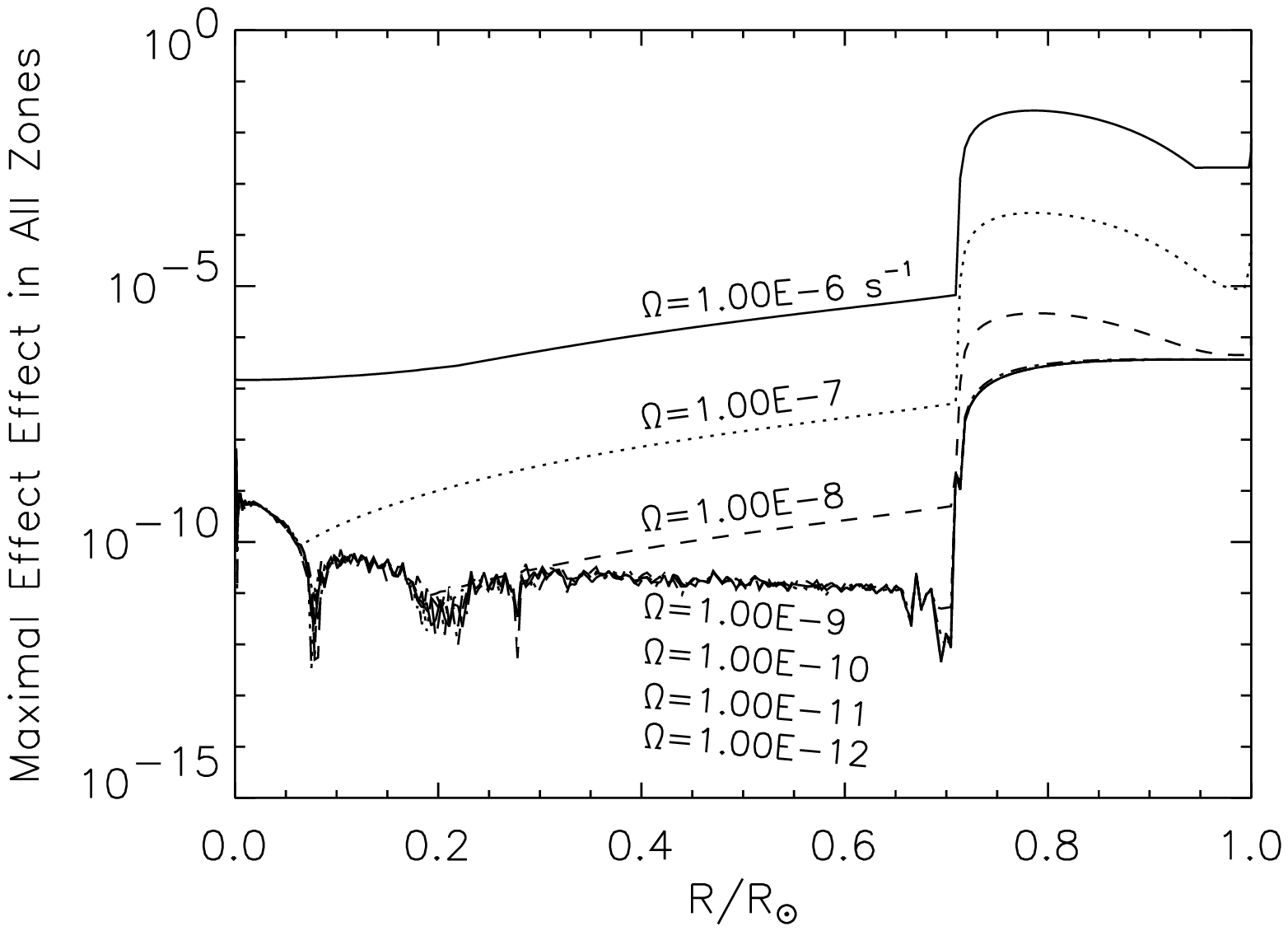}
\caption{Maximal rotation effect in all zones as a function of both the rotation rate $\Omega$
and radius $R/R_\sun$.}\label{fig:eps4}
\end{figure}

\begin{figure}
\plotone{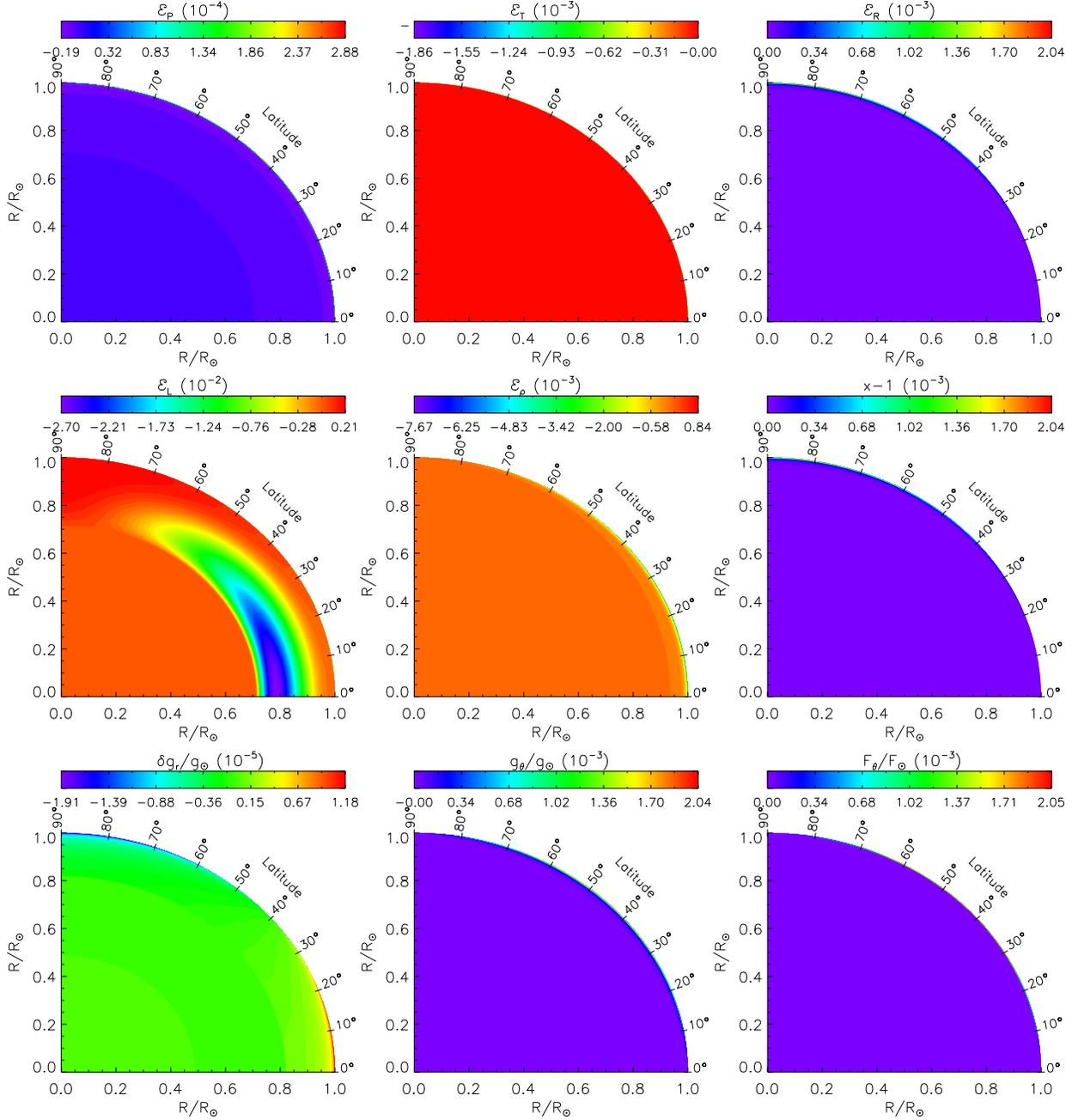}
\caption{Contours of the 2D solar model with a uniform rotation-like toroidal
magnetic field $\vb{B}=(0,0,(4\pi\rho)^{1/2}\Omega r\sin\theta)$, 
where $\Omega=10^{-6}$ s$^{-1}$. The top five sub-figures show the detail dependence of $\h{E}_P$
to $\h{E}_\rho$ on $R/R_\sun$ and $\theta$. The last fore sub-figures show the equipotential 
surface function $x-1$, the transverse component of energy flux $\vb{F}$, $F_\theta/F_\sun$, 
the radial perturbation component, and transverse component of the gravitational acceleration.}
\label{fig:roth}
\end{figure}

The uniform rotation-like toroidal magnetic field is $\vb{B}=(0,0,(4\pi\rho)^{1/2}\Omega
r\sin\theta)$. We repeat the similar model calculations to rotation. Figs.~\ref{fig:eps3}
-\ref{fig:roth} show the results. Once again, the high-precision is achieved. 
Comparing them with Figs.\ref{fig:eps1}-\ref{fig:rot} we 
can see rotation-like magnetic fields affect stellar structures in a different way 
from the rotation: magnetic effects take place in the convection zone and peak 
near the surface. Rotation-like turbulence behaves like a rotation-like magnetic field.

\subsubsection{Differential rotation-like magnetic field: torus}

The torus field is a rotation-like toroidal magnetic field,
$\vb{B}=(0,0,(4\pi\rho)^{1/2}\Omega r\sin\theta)$. The magnetic rotation rate $\Omega$
is defined in Appendix~\ref{app:torus}. There are two torus tubes that are parallel to 
the equatorial plane since they are assumed to be symmetric with respect 
to the equatorial plane. As a result, there are four circles on any meridional plane.

Unlike the uniform rotation rate, we should first find out the discrete Fourier transform
of the square of the differential rotation rate $\Omega$, $\Omega^2$, which is equally 
discretized in the range of $\theta$ from 0 to $\pi/2$, namely $\Omega_i$ for i = 0 to N. 
Here N should be a power of 2. 
We calculate $\Omega^2$ in the first quadrant and then extend it to the other three quadrants
according to the symmetry described above. Its discrete Fourier transform $F_n$ are
finally calculated by means of the Fast Fourier Transform (FFT) of a real function (See the
subroutine realft.for given in Numerical Recipe) for n = 0 to 4N. Each pair of the data contain 
the real and imaginary parts of the FFT except for the first pair. The imaginary part vanishes
since $\Omega^2$ is a real function of $\theta$, which is now in the 
range of 0 to $2\pi$. The odd components vanish due the equatorial symmetry. We use $y_n$ to denote
the nonzero components. The nonzero $F_n$ contains $F_0$, which is twice the uniform component,
$y_0=F_0/2$; and $F_1$, which stores the twice of the Nyquist critical wavenumber component, 
$y_N = F_1/2$; and the even components $y_n=F_{4n}$ for n = 1 to N-1. Consequently, we have
\be
  \Omega^2 = \sum_{n=0}^N y_n \cos2n\theta.
\ee

Fig.~\ref{fig:torus1} contains nine sub-figures for the Gaussian profile defined 
in Appendix B, in which $\Omega_0=3\times10^{-5}$. Sub-figure (1,1) shows the reciprocal of 
the plasma $\beta$ parameter as a function of ($R/R_\sun,\theta$), 
which is defined as the ratio of the gas pressure over the 
magnetic pressure: $1/\beta=\frac{1}{2}\rho\Omega^2r^2\sin^2\theta/P$.
Sub-figures (1,2)-(2,3) show $\h{E}_P\sim \h{E}_\rho$. The equipotential surface, 
the colatitudinal components of the gravitational 
acceleration vector and the flux vector are shown in the bottom panel, namely, 
sub-figures (3,1)-(3,3).

Sub-figure (1,2) shows that pressure does not vary with colatitude $\theta$ 
on the equipotential surface. It is the very feature that is required by the hydrostatic 
equilibrium on the surface. The numerical method of the solution to the 2D stellar 
structure equations presented in this paper is designed to achieve this feature. It is not 
trivial at all.

Sub-figure (1,3) indicates that the presence of the magnetic flux loop beneath the surface
affects the temperature distribution in site and above. This is reasonable since the 
thermal time scale near the base of the convection zone (where the loop is located) 
is much longer than the solar cycle so that the temperature perturbation travels little 
inwards in the cyclic period. In contrast, it can substantially travel outwards in short 
time since the thermal timescale above the torus field is very small. Another feature 
for the 2D temperature effect is that the temperature increases above the buried field.
We see sunspots in the solar active regions. It is well-known that sunspots reduce the 
energy output of the Sun. We also know that the active regions increase the net energy 
output of the Sun as a whole. The idea that the temperature increase caused by the buried fields 
over-compensates the sunspot is a natural explanation to the net increase of 
the energy output in the active regions of the Sun.

Sub-figures (2,1) and (3,1) are similar to each other. The distinction is their references: 
the former refers to the 1D radius of the equipotential surface, and the latter refers 
to the equatorial radius. The maximal radius change takes place at the minimal 
$\beta$ parameter. Both of them show the equipotential surface profile.

Comparing sub-figure (2,3) with (1,1) we can see that the density 
change inversely follows the plasma $\beta$ parameter and is of the same order of 
magnitude as $1/\beta$, which is in agreement with the analytical result: 
$(\rho\, - <\rho>)/<\rho>=1/(1+1/\beta)\approx - 1/\beta$. 
The sub-figure also shows that the density decrease maximizes in the loop. This will give 
rise to a buoyant force on the loop in the radial direction. Its component on the plane 
that is parallel to the equator plane cancels out since the loop is azimuthally symmetric. 
Its component in the meridional direction will generate an acceleration in the same 
direction, $a_m$. Detailed calculation (see Appendix \S\ref{app:torus}) 
shows $a_m\approx 32$ cm s$^{-2}$. 
The buoyant force is assumed to be balanced by the turbulent pressure 
generated by the down-flow plumes found in the realistic three-dimensional turbulent 
simulations of the solar convection zone near the surface of the Sun (e.g., Stein and 
Nordlund 1998; Robinson et al 2003). These simulations reveal 
that the up-flow and down-flow are not symmetric and the down-flow 
is stronger than the up-flow. 

In the real Sun, this condition is obeyed until the magnetic 
field reaches a critical value whereby the buoyancy forces dominate,
magnetic loops making up the torus float up, produce magnetic 
activity in the solar surface, and the toroidal field is depleted. 
We do not model these details in our code excepting in terms of the 
decrease of the toroidal field.

The transverse components of the gravitational acceleration vector $\vb{g}$ 
and the flux $\vb{F}$ shown in sub-figures (3,2) and (3,3) are 
purely 2D effects. Their characteristics and other 2D effects 
need to be investigated further and will be presented separately.

\section{CONCLUSIONS}

We present a new set of differential equations to describe the stellar
equilibrium, in which two dimensional effects are explicitly taken into account.
We improve the treatment presented in a previous paper of this series, by
relaxing some approximations that had been made in that context; this task 
required one more differential equation, with the introduction of a new 
variable, i.e. the deviation of the radial component of gravity from the 
standard expression that is obtained when the Poisson equation is solved 
neglecting the angular derivatives.

We have shown that by selecting an appropriate convergence criterion our code can reach 
the precision required by current and forthcoming observations.

The code can now be used to test the effects of magnetic fields of
any axisymmetric magnetic field configuration on the structure of the 
current Sun, and to investigate the change of the observable solar 
properties related to the variation of the magnetic field with the solar cycle.
We have used the code to scan a very large region of the parameter space 
to test the code, and will present our findings in a separate paper.

Finally, we wish to emphasize that because we are interested 
in modeling the effects of a dynamo-type 
field on the detailed envelope structure and global properties of 
the Sun, the code has been optimized for short timescales phenomena 
(down to 1 yr).  Consequently, the time dependence of the code has so far
been tested exclusively to address such problems, and we can not 
assume that the code could be used to 
model long term stellar evolution without further modifications.

\acknowledgements 
We want to acknowledge the following support for this work: LLH by NSF Grant 
ATM 073770, and the Vetlesen Foundation; SS by the Vetlesen and the Brinson 
Foundations; SB by NSF grants ATM 0348837 and ATM 0737770; SLB by MSTC grant 
2007CB815406, NSFC grants 10433030, 10773003, 10778601, and PD by NASA 
grant NAG5-13299.

\appendix

\section{COEFFICIENT MATRIX ELEMENTS}\label{app:coeff}

Eq.~(\ref{eq:linear}) consists of a set of non-homogeneous linear 
algebraic equations. We work out these nonzero elements in this appendix. 

\subsection{Useful Partial Derivatives}\label{sct:upd}

The partial derivatives of the differential equations are 
required for the linearization. By defining the shorthand 
notation $\partial_X Y=\partial Y/\partial\log X$, we can 
calculate the useful derivatives as follows.

In fact, we need to calculate all the derivatives of 
$\h{P}$, $\h{T}$,  $\h{U}^i$ ($i = 1, 2, 3, 4, 5$), 
$\h{R}$, and $\h{L}^i$ ($i = 1, 2$) with respect to $P'$, 
$T'$, $r'$, $L$, and $U$, respectively. For the sake 
of completeness and conciseness, we write down all 
nonzero partial derivatives and formulas  except for 
the same as in Paper I. The derivatives of $\h{P}$, $\h{T}$, and
$\h{L}^1$ are the same as in Paper I, where $\h{L}^1$ 
is equivalent to $\h{L}$ in Paper I.

The derivatives of $\h{R}$ may be nonzero only for k = j and l = i - 1, i. 
The unique nonzero derivative is
\[
\partial_{R}\h{R}=-\cdot\h{R},
\]
which is different from Paper I.

The derivatives of $\h{L}^\ell$ ($\ell=2,3$) may be nonzero not only for
k = j and l = i - 1, i. For the sake of simplicity, 
we rewrite $F_\theta$ as follows:
\[
  F_\theta = (F^1 + F^2 + F^3 )\tilde{P},
\]
where
\Ba
 F^1  &=&  -\frac{4ac T^4\nabla}{3\kappa\rho}    \\
 F^2  &=&  -\frac{1}{2} \frac{\rho C_PT l_m v_{\s{conv}}\nabla}
   {1+v_{\s{conv}}/v_0}  \\
 F^3 &=& \frac{1}{2} \frac{\rho C_PT l_m v_{\s{conv}}\nabla'_{\s{ad}}}
   {1+v_{\s{conv}}/v_0} \\
 \tilde{P} &=& \frac{Gm Q \Lambda}{4\pi r_e^4 P_T}.  
\Ea
In order to obtain the nonzero derivatives of $F_\theta$, we 
also need the following formulas:
\Ba
 \partial_P F^1 &=& -F^1(\kappa_P +\alpha - \nabla_P) \\
 \partial_T F^1 &=& -F^1(\kappa_T -\delta - \nabla_T-4) \\
 \partial_R F^1 &=& F^1\nabla_R \\
 \partial_L F^1 &=& F^1 \nabla_L \\
 \partial_P F^2 &=& F^2(\alpha + \nabla_P + C_{\s{PP}}) 
    -\frac{v_{\s{conv}}/v_0}{1+v_{\s{conv}}/v_0}
   F^2(2\alpha+C_{\s{PP}}+\kappa_P) \\
 \partial_T F^2 &=& F^2(1-\delta + \nabla_T + C_{\s{PT}}) 
    -\frac{v_{\s{conv}}/v_0}{1+v_{\s{conv}}/v_0}
   F^2(-2\delta+C_{\s{PT}}+\kappa_T-3) \\
 \partial_R F^2 &=& F^2\nabla_R \\
 \partial_L F^2 &=& F^2\nabla_L \\
 \partial_P F^3 &=& F^3(\alpha + \nabla'_P + C_{\s{PP}}) 
    -\frac{v_{\s{conv}}/v_0}{1+v_{\s{conv}}/v_0}
   F^3(2\alpha+C_{\s{PP}}+\kappa_P) \\
 \partial_T F^3 &=& F^3(1-\delta + \nabla'_T + C_{\s{PT}}) 
    -\frac{v_{\s{conv}}/v_0}{1+v_{\s{conv}}/v_0}
   F^3(-2\delta+C_{\s{PT}}+\kappa_T-3) \\
 \partial_R F^3 &=& 0 \\
 \partial_P \tilde{P} &=& -\tilde{P} \\
 \partial_T \tilde{P} &=& 0 \\
 \partial_R \tilde{P} &=& -4\tilde{P}
\Ea
Here $\kappa_P\equiv\left(\p{\ln\kappa}{\ln P_T}\right)_T$, $\kappa_T\equiv\left(\p{\ln\kappa}{\ln T}\right)_{P_T}$, 
$C_{\s{PP}}\equiv\left(\p{\ln C_P}{\ln P_T}\right)_T$, $C_{\s{PT}}\equiv\left(\p{\ln C_P}{\ln T}\right)_{P_T}$, 
and $v_0=6acT^3/\rho^2 C_P l_m\kappa$, $\nabla'_P=(\partial\ln\nabla'_{\s{ad}}/\partial\ln P_T)_T$, and $\nabla'_T=(\partial\ln\nabla'_{\s{ad}}/\partial\ln T)_{P_T}$.
As a result, we have
\Ba
 \partial_P F_\theta &=& \tilde{P}\sum_{\ell=1}^3 \partial_P F^\ell
   + \partial_P\tilde{P}\sum_{\ell=1}^3 F^\ell \\
 \partial_T F_\theta &=& \tilde{P}\sum_{\ell=1}^3 \partial_T F^\ell
   + \partial_T\tilde{P}\sum_{\ell=1}^3 F^\ell \\
 \partial_R F_\theta &=& \tilde{P}\sum_{\ell=1}^2 \partial_R F^\ell
   + \partial_R\tilde{P}\sum_{\ell=1}^3 F^\ell \\
 \partial_L F_\theta &=& \tilde{P}\sum_{\ell=1}^2 \partial_L F^\ell \\
 \h{F} &=& \partial_PF_\theta + \partial_T F_\theta\cdot\nabla 
   + \partial_R F_\theta\cdot\p{r'}{P'} + \partial_L F_\theta\cdot\h{L}/\h{P}
\Ea
where
\[
\p{r'}{P'} = -\frac{4\pi r_e^3P_T}{G m Q x}.
\]
These finish the expressions for $\h{L}^2$ and $\h{L}^3$, and their derivatives:
\Ba
\partial_P\h{L}^2 &=& \h{L}^2(F^{-1}_\theta \partial_P F_\theta-\alpha) \\
\partial_T\h{L}^2 &=& \h{L}^2(F^{-1}_\theta \partial_T F_\theta+\delta) \\
\partial_R\h{L}^2 &=& \h{L}^2(F^{-1}_\theta \partial_R F_\theta-1) \\
\partial_L\h{L}^2 &=& \h{L}^2 F^{-1}_\theta \partial_L F_\theta \\
\partial_P\h{L}^3 &=& -\alpha\cdot\h{L}^3 \\
\partial_T\h{L}^3 &=& \delta\cdot\h{L}^3 \\
\partial_R\h{L}^3 &=& -\h{L}^3
\Ea 

After all nonzero components and their derivatives are 
calculated, we can sum them to obtain
\Ba
\h{L} &=& \sum_{\ell=1}^3\h{L}^\ell \\
\partial_P\h{L} &=& \sum_{\ell=1}^3\partial_P\h{L}^\ell \\
\partial_T\h{L} &=& \sum_{\ell=1}^3\partial_T\h{L}^\ell \\
\partial_R\h{L} &=& \sum_{\ell=1}^3\partial_R\h{L}^\ell \\
\partial_L\h{L} &=& \partial_L\h{L}^2
\Ea

\subsection{INTERIOR POINTS}\label{sct:a2}

The interior points can be grouped into four blocks:
\begin{description}
  \item{Block I,} l = i - 1 and k = j,
  \item{Block II,} l = i and k = j.
\end{description}

\subsubsection{w = P}

For block I,
\Ba
\p{F^{i}_P}{R'_{i-1}} &=& -\frac{1}{2}\Delta s_i\partial_R\h{P}_{i-1} \\
\p{F^{i}_P}{L_{i-1j}} &=& 0 \\
\p{F^{i}_P}{P'_{i-1}} &=& -\frac{1}{2}\Delta s_i\partial_P\h{P}_{i-1} -1 \\
\p{F^{i}_P}{T'_{i-1j}} &=& 0 \\
\Ea
For block II,
\Ba
\p{F^{i}_P}{R'_{i}} &=& -\frac{1}{2}\Delta s_i\partial_R\h{P}_{i} \\
\p{F^{i}_P}{L_{ij}} &=& 0 \\
\p{F^{i}_P}{P'_{i}} &=& -\frac{1}{2}\Delta s_i\partial_P\h{P}_{i} + 1 \\
\p{F^{i}_P}{T'_{ij}} &=& 0 \\
\Ea

\subsubsection{w = T}

For block I,
\Ba
\p{F^{ij}_T}{R'_{i-1}} &=& -\frac{1}{2}\Delta s_i\partial_R\h{T}_{i-1j} \\
\p{F^{ij}_T}{L_{i-1j}} &=& -\frac{1}{2}\Delta s_i\partial_L\h{T}_{i-1j} \\
\p{F^{ij}_T}{P'_{i-1}} &=& -\frac{1}{2}\Delta s_i\partial_P\h{T}_{i-1j} \\
\p{F^{ij}_T}{T'_{i-1j}} &=& -\frac{1}{2}\Delta s_i\partial_T\h{T}_{i-1j} -1 
\Ea
For block II,
\Ba
\p{F^{ij}_T}{R'_{i}} &=& -\frac{1}{2}\Delta s_i\partial_R\h{T}_{ij} \\
\p{F^{ij}_T}{L_{ij}} &=& -\frac{1}{2}\Delta s_i\partial_L\h{T}_{ij} \\
\p{F^{ij}_T}{P'_{i}} &=& -\frac{1}{2}\Delta s_i\partial_P\h{T}_{ij} \\
\p{F^{ij}_T}{T'_{ij}} &=& -\frac{1}{2}\Delta s_i\partial_P\h{T}_{ij} + 1 
\Ea

\subsubsection{w=R}

For block I,
\begin{eqnarray*}
  \p{F^{i}_R}{R'_{i-1}} &=& -\frac{1}{2}\Delta s_i \partial_R\h{R}_{i-1}-1 \\ 
  \p{F^{i}_R}{L_{i-1j}} &=&  0 \\ 
  \p{F^{i}_R}{U_{i-1j}} &=&  0 \\ 
  \p{F^{i}_R}{P'_{i-1}} &=& -\frac{1}{2}\Delta s_i \partial_P\h{R}_{i-1} \\ 
  \p{F^{i}_R}{T'_{i-1j}} &=& -\frac{1}{2}\Delta s_i \partial_T\h{R}_{i-1} \\
\end{eqnarray*}
For block II,
\begin{eqnarray*}
  \p{F^{i}_R}{R'_{i}} &=& -\frac{1}{2}\Delta s_i \partial_R\h{R}_{i} +1 \\
  \p{F^{i}_R}{L_{ij}} &=&  0 \\
  \p{F^{i}_R}{U_{ij}} &=&  0 \\
  \p{F^{i}_R}{P'_{i}} &=& -\frac{1}{2}\Delta s_i \partial_P\h{R}_{i} \\
  \p{F^{i}_R}{T'_{ij}} &=& -\frac{1}{2}\Delta s_i \partial_T\h{R}_{i}
\end{eqnarray*}

\subsubsection{w = L}

For block I,
\Ba
\p{F^{ij}_L}{R'_{i-1}} &=& -\frac{1}{2}\Delta s_i\partial_R\h{L}_{i-1j} \\
\p{F^{ij}_L}{L_{i-1j}} &=& 0 \\
\p{F^{ij}_L}{U_{i-1j}} &=& 0  \\
\p{F^{ij}_L}{P'_{i-1}} &=& \frac{1}{2}\Delta s_i\partial_P\h{L}_{i-1j} \\
\p{F^{ij}_L}{T'_{i-1j}} &=& -\frac{1}{2}\Delta s_i\partial_T\h{L}_{i-1j} -1 
\Ea
For block II,
\Ba
\p{F^{ij}_L}{R'_{i}} &=& -\frac{1}{2}\Delta s_i\partial_R\h{L}_{ij} \\
\p{F^{ij}_L}{L_{ij}} &=& 0 \\
\p{F^{ij}_L}{U_{ij}} &=& 0 \\
\p{F^{ij}_L}{P'_{i}} &=& -\frac{1}{2}\Delta s_i\partial_P\h{L}_{ij} \\
\p{F^{ij}_L}{T'_{ij}} &=& -\frac{1}{2}\Delta s_i\partial_P\h{L}_{ij} + 1 
\Ea

\subsection{BOUNDARY POINTS}\label{sct:a3}

\subsubsection{Center: w = R}

Central boundary points have only block II for w = R:
\begin{eqnarray*}
  \p{F^{1}_R}{R'_{1}} &=& 1 \\
  \p{F^{1}_R}{L_{1j}} &=&  0 \\ 
  \p{F^{1}_R}{U_{1j}} &=&  0 \\ 
  \p{F^{1}_R}{P'_{1}} &=& \frac{1}{3}\alpha_{01} \\
  \p{F^{1}_R}{T'_{1j}} &=& -\frac{1}{3}\delta_{01}
\end{eqnarray*}

\subsubsection{Center: w = L}

Central boundary points have block II for w = L:
\begin{eqnarray*}
\p{F^{1j}_L}{R'_{1}} &=& 0 \\
\p{F^{1j}_L}{L_{1j}} &=&  1 \\
\p{F^{1j}_L}{U_{1j}} &=&  0 \\
\p{F^{1j}_L}{P'_{1}} &=& -\partial_P\h{L}_{1j} \\
\p{F^{1j}_L}{T'_{1j}} &=& -\partial_T\h{L}_{1j} 
\end{eqnarray*}

\subsubsection{Surface: w = R}

Surface boundary points have block I for w = R:
\begin{eqnarray*}
\p{F^{M+1}_R}{R'_{M}} &=& 1 \\
\p{F^{M+1}_R}{L_{Mj}} &=&  0 \\
\p{F^{M+1}_R}{U_{Mj}} &=&  0 \\
\p{F^{M+1}_R}{P'_{Mj}} &=& -a_1 \\
\p{F^{M+1}_R}{T'_{Mj}} &=& -a_2 
\end{eqnarray*}

\subsubsection{Surface: w = L}

Surface boundary points have block I for w = L:
\begin{eqnarray*}
\p{F^{M+1j}_L}{R'_{Mj}} &=& 0\\
\p{F^{M+1j}_L}{L_{Mj}} &=&  1 \\
\p{F^{M+1j}_L}{U_{Mj}} &=&  0 \\
\p{F^{M+1j}_L}{P'_{Mj}} &=& -L_{Mj}a_4 \\
\p{F^{M+1j}_L}{T'_{Mj}} &=& -L_{Mj}a_5 
\end{eqnarray*}

\section{Buoyant acceleration of a magnetic flux loop in the meridional direction}\label{app:torus}

The magnetic flux loop used in this paper is assumed to be axisymmetric with 
respect to the polar axis. Its buoyant force ($\vb{f}_B$) is radial and can 
be decomposed into two components. One is parallel to the equatorial plane ($f_e$), 
and the other is perpendicular to it ($f_m$). The former is canceled 
out since the loop is axisymmetric with respect to the polar axis (i.e., $f_e=0$),
and the latter is in the meridional direction ($f_m\ne 0$). In order 
to compute the buoyant acceleration of the loop in the meridional direction 
($a_m=f_m/m_L$), we have to compute $f_m$ and the mass of the loop $m_L$.

We must first calculate the boundary of the loop. The polar axis is assumed 
to be the z-axis. The equation for a torus azimuthally 
symmetric about the z-axis in Cartesian coordinates is
\be
  (c-\sqrt{x^2+y^2})^2 + (z-z_0)^2 = a^2,
\ee
where $c$ is the radius from the center of the hole to the center of the 
torus tube, $a$ is the radius of the tube, and $(0,0,z_0)$ is the center 
point coordinate of the hole. In the xz-plane the torus becomes two circles. 
One of them is
\be
  (c-x)^2+(z-z_0)^2=a^2 \label{eq:xz1}
\ee
in Cartesian coordinates. We need to determine its boundary.
In the spherical polar coordinates ($r,\theta,\phi$),  Eq.~(\ref{eq:xz1}) becomes
\be
  (c-r\sin\theta)^2 + (r\cos\theta-c\cot\theta_0)^2 = a^2, \label{eq:ra1}
\ee
where $\theta_0$ is the colatitude of the center of the circle.
The radius range of the circle for each $\theta$ is given by the solutions for $r$ of  
Eq.~(\ref{eq:ra1}): $r_-\le r\le r++$, where $r_\pm$ are defined by
\begin{equation}
  r_\pm=c(\sin\theta+\cos\theta\cot\theta_0)\pm c[(\sin\theta
+\cos\theta\cot\theta_0)^2-1-\cot^2\theta_0+a^2/c^2]^{1/2}. \label{eq:rpm}
\end{equation}
The colatitude range of the circle for each radius $r$ is determined by the solutions 
of Eq.~(\ref{eq:ra1}) for $\theta$:
\be
  \theta_\pm = \arccos\left[\frac{b\sin2\theta_0\pm[b^2/\sin^22\theta_0-4(b^2-1)\sin^2\theta_0]^{1/2}}{2}\right],
\ee
where
\be
 b=\frac{c^2/\sin^2\theta_0+r^2-a^2}{2cr}.
\ee
Since $\theta_-\ge\theta_+$, the boundary of Eq.~(\ref{eq:xz1}) can be expressed by
\be
\h{C}:\hspace{4mm}  r_- \le r \le r_+, \mbox{ and } \theta_+\le \theta \le \theta_-, 
\mbox{ and } 0\le\phi\le2\pi.
\ee

We have two ways to define a torus field. One is to use the step function: 
$\Omega=\Omega_0$ within the loop confined by $\h{C}$, but $\Omega=0$ 
outside the loop, where $\Omega_0$ is a constant. The other way is to use the 
Gaussian profile to smooth the step function: 
$\Omega=\Omega_0\exp[-\frac{1}{2}\frac{(\theta-\theta_0)^2}{\sigma^2}]$,
where $\sigma=\frac{1}{3}(\theta_+-\theta_-)$.

We can then express the meridional buoyant force component $f_m$ and mass 
in the loop in terms of the following integrals:
\ba
  f_m &=& 2\pi c\cos\theta_0\int_{\s{\h{C}}} r g(<\rho>-\rho)dr d\theta, \\
  m_L &=& 2\pi c\int_{\s{\h{C}}} r\rho dr d\theta.
\ea
The acceleration equals $a_B=f_m/m_L$. Here $<\rho>$ is the averaged density over 
the colatitude $\theta$ from 0 to $\pi/2$.

{}

\clearpage

\clearpage

\clearpage

\clearpage

\clearpage

\clearpage

\clearpage

\clearpage
\begin{figure}
\plotone{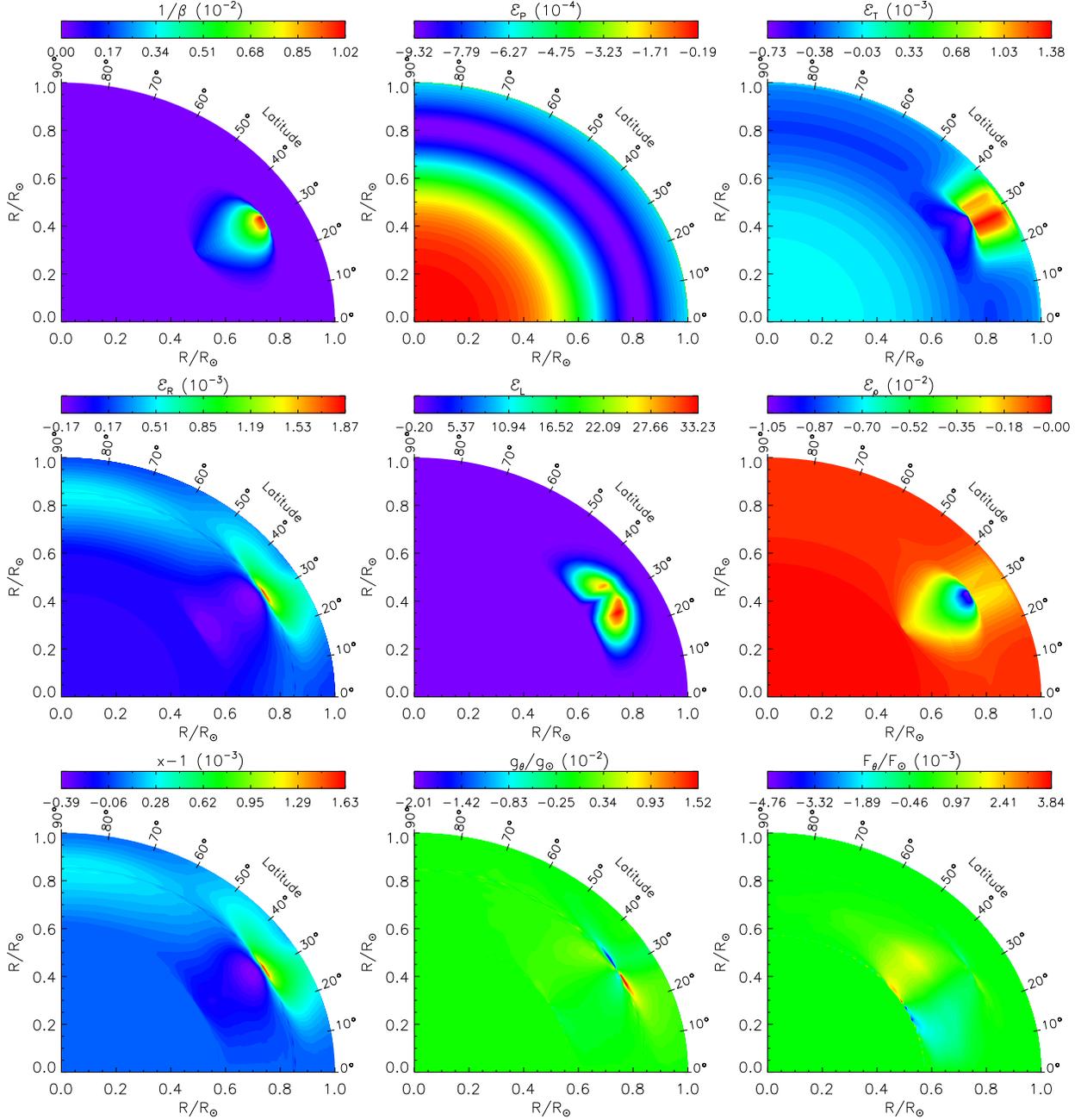}
\caption{Contours of the 2D solar variability model with a torus field, 
in which the applied magnetic field (measured in the plasma $\beta$ parameter), the relative changes 
of the stellar structure variables (pressure, temperature, radius, luminosity and density)
and the transverse components of the gravitational acceleration and flux vectors.}\label{fig:torus1}
\end{figure}

\end{document}